\documentclass{emulateapj}
\usepackage{graphicx}
\usepackage{natbib}
\newcommand{\be}{\begin{equation}}

\newcommand{\ee}{\end{equation}}
\newcommand{\ba}{\begin{eqnarray}}
\newcommand{\ea}{\end{eqnarray}}
\newcommand\bp{\begin{figure}}
\newcommand\ep{\end{figure}}
\newcommand\bpm{\begin{figure*}}
\newcommand\epm{\end{figure*}}
\newcommand{\btab}{\begin{tabular}}
\newcommand{\etab}{\end{tabular}}
\newcommand{\bt}{\begin{table}}
\newcommand{\et}{\end{table}}
\newcommand{\ben}{\begin{enumerate}}
\newcommand{\een}{\end{enumerate}}
\newcommand\reffig[1]{Figure \ref{fig:#1}}

\newcommand\refsec[1]{\S \ref{sec:#1}}
\newcommand\reftbl[1]{Table \ref{tbl:#1}}
\newcommand\refapp[1]{Appendix \ref{app:#1}}

\newcommand{\GeV}{\rm{GeV}}

\newcommand\mean[1]{\langle #1 \rangle}

\newcommand{\Fermi}{\emph{Fermi}}

\newcommand{\bcn}{\begin{center}}
\newcommand{\ecn}{\end{center}}

\newcommand{\degree}{^\circ}
\newcommand{\epp}{e^+e^-}

\begin{document}

\title{The \emph{Fermi} Haze: \\
       A Gamma-Ray Counterpart to the Microwave Haze}

\author{Gregory Dobler,\altaffilmark{1,2,5}
  Douglas P. Finkbeiner,\altaffilmark{1,3} 
  Ilias Cholis,\altaffilmark{4} \\
  Tracy Slatyer,\altaffilmark{1,3}
  \& Neal Weiner\altaffilmark{4}}

\altaffiltext{1}{ 
  Institute for Theory and Computation,
  Harvard-Smithsonian Center for Astrophysics, 
  60 Garden Street, MS-51, Cambridge, MA 02138 USA } 
\altaffiltext{2}{
  Kavli Institute for Theoretical Physics, 
  University of California, Santa Barbara
  Kohn Hall, Santa Barbara, CA 93106 USA
}
\altaffiltext{3}{ 
  Physics Department, 
  Harvard University, 
  Cambridge, MA 02138 USA }
\altaffiltext{4}{ 
  Center for Cosmology and Particle Physics, 
  Department of Physics, 
  New York University, New York, NY 10003 USA }
\altaffiltext{5}{dobler@kitp.ucsb.edu}

\begin{abstract}
  The \emph{Fermi Gamma-Ray Space Telescope} reveals a diffuse inverse
  Compton signal in the inner Galaxy with a similar spatial morphology
  to the microwave haze observed by WMAP, supporting the synchrotron
  interpretation of the microwave signal.  Using spatial templates, we
  regress out $\pi^0$ gammas, as well as IC and bremsstrahlung
  components associated with known soft-synchrotron counterparts.  We
  find a significant gamma-ray excess towards the Galactic center with
  a spectrum that is significantly harder than other sky components
  and is most consistent with IC from a hard population of electrons.
  The morphology and spectrum are consistent with it being the IC
  counterpart to the electrons which generate the microwave haze seen
  at WMAP frequencies.  In addition, the implied electron spectrum is hard;
  electrons accelerated in supernova shocks in the disk which then
  diffuse a few kpc to the haze region would have a softer spectrum.
  We describe the full sky \emph{Fermi} maps used in this analysis and
  make them available for download.
\end{abstract}

\section{Introduction}

The most detailed and sensitive maps of diffuse microwave emission in
our Galaxy have been produced by the \emph{Wilkinson Microwave
  Anisotropy Probe} (WMAP).  An analysis of the different emission
mechanisms in these maps uncovered a microwave ``haze'' towards the
Galactic center (GC) that has roughly spherical morphology and radius
$\sim4$ kpc \citep{Finkbeiner:2003im}.  Since its discovery,
\cite{Finkbeiner:2004us} and \cite{Dobler:2008ww} have argued that the
microwave haze is hard synchrotron emission due to the fact that
alternative hypotheses such as free-free (thermal bremsstrahlung of
the ionized gas), spinning dust, or thermal dust have difficulty
explaining its morphology, spectrum, or both.

However, the 23-33 GHz spectrum of the haze synchrotron is harder than
that expected from diffused electrons originally accelerated by
supernova shocks in the plane \citep{Dobler:2008ww}.  Some variation is expected in
the synchrotron spectrum, but generally in the sense that it should be
\emph{softer} at higher frequencies since the electrons lose energy
preferentially at high energies as they diffuse from their source.
While there are significant uncertainties in the spectrum of the haze
\citep[see the discussion in][]{Dobler:2008ww}, the data require that
the diffused spectrum be roughly as hard as the expected
\emph{injection} spectrum from first-order Fermi acceleration at
supernova shock fronts (number density $dN/dE \propto E^{-2}$).  That
is, if the electrons were produced in shocks in the disk, then they would have to
have undergone no diffusive energy losses over a $\sim4\pi/3 \ (4$
kpc$)^3$ volume, which seems unlikely.  Furthermore, there is
significant emission in WMAP 23, 33, and 41 GHz bands from electrons
that \emph{were} generated in SN shocks; this emission has a very
disk-like morphology (and softer spectral index which is consistent with
shock acceleration), while the haze has a more spherical
morphology.\footnote{Hereafter ``spherical'' is taken to mean ``not
  disk-like'' -- if anything, the haze is non-spherical in the direction
  perpendicular to the disk (see \refsec{residualmaps}).}

Together, the haze spectrum and morphology imply either (1) a new class of
objects distributed in the Galactic bulge and largely missing from the
disk; (2) significant acceleration from shocks several kpc off the
plane towards the Galactic center; or, perhaps most intriguingly, (3) a new electron
component from a new physical mechanism.
The claim that novel physics or astrophysics is required to
explain the WMAP data is called the \emph{haze hypothesis} to
distinguish it from the two null hypotheses: (1) that the microwave haze is not
synchrotron, but rather some combination of free-free and spinning
dust; and (2) that the haze is synchrotron, but the electron spectrum
required is not unusual.  In this work we do not address the origin of
the electrons, but instead consider what the data from the \emph{Fermi
  Gamma-ray Space Telescope}\footnote{See
  \texttt{http://fermi.gsfc.nasa.gov/ssc/data/}} imply for their
existence.

\bpm
  \includegraphics[width=0.49\textwidth]{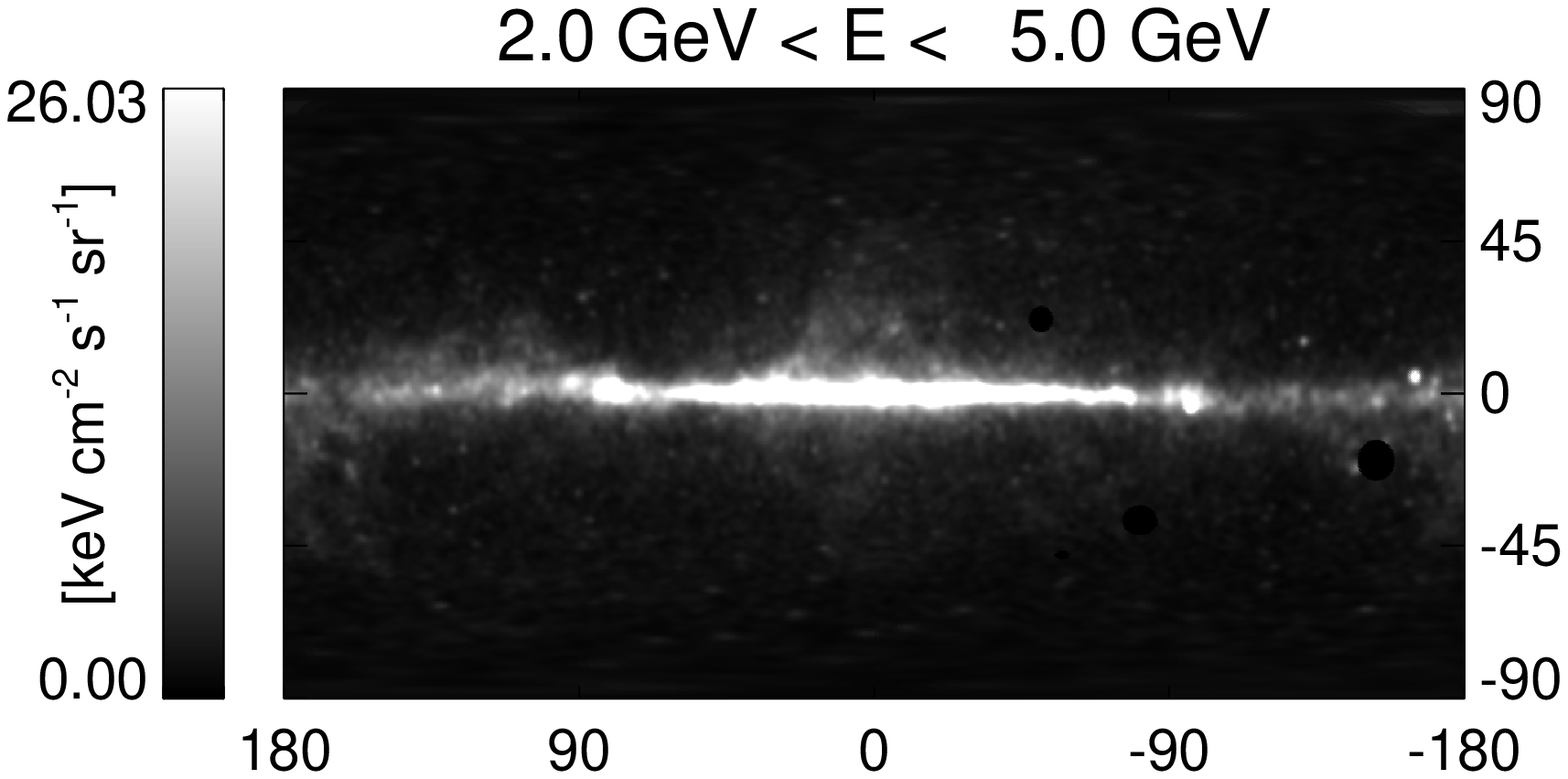}
  \includegraphics[width=0.49\textwidth]{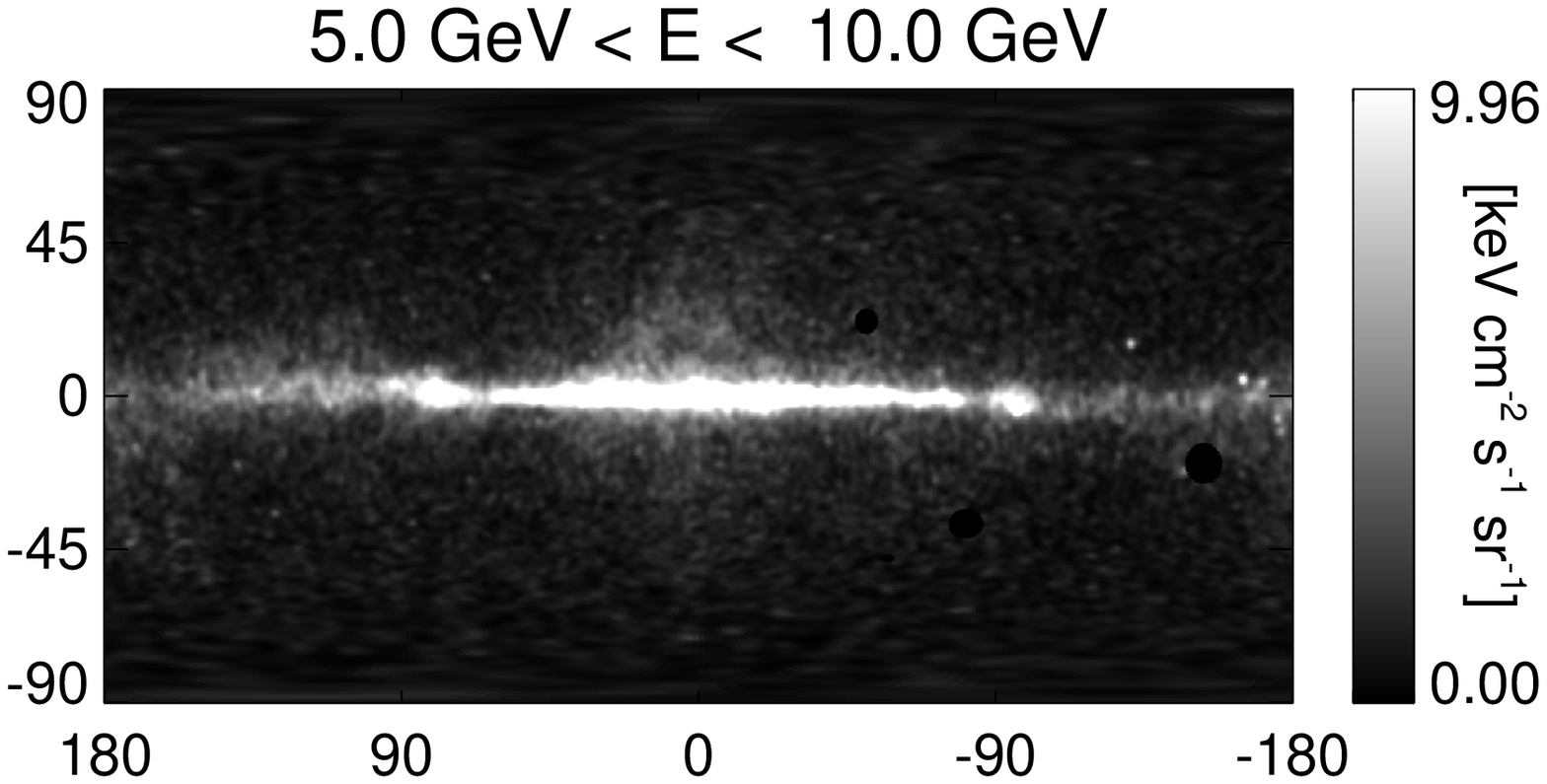}

  \includegraphics[width=0.49\textwidth]{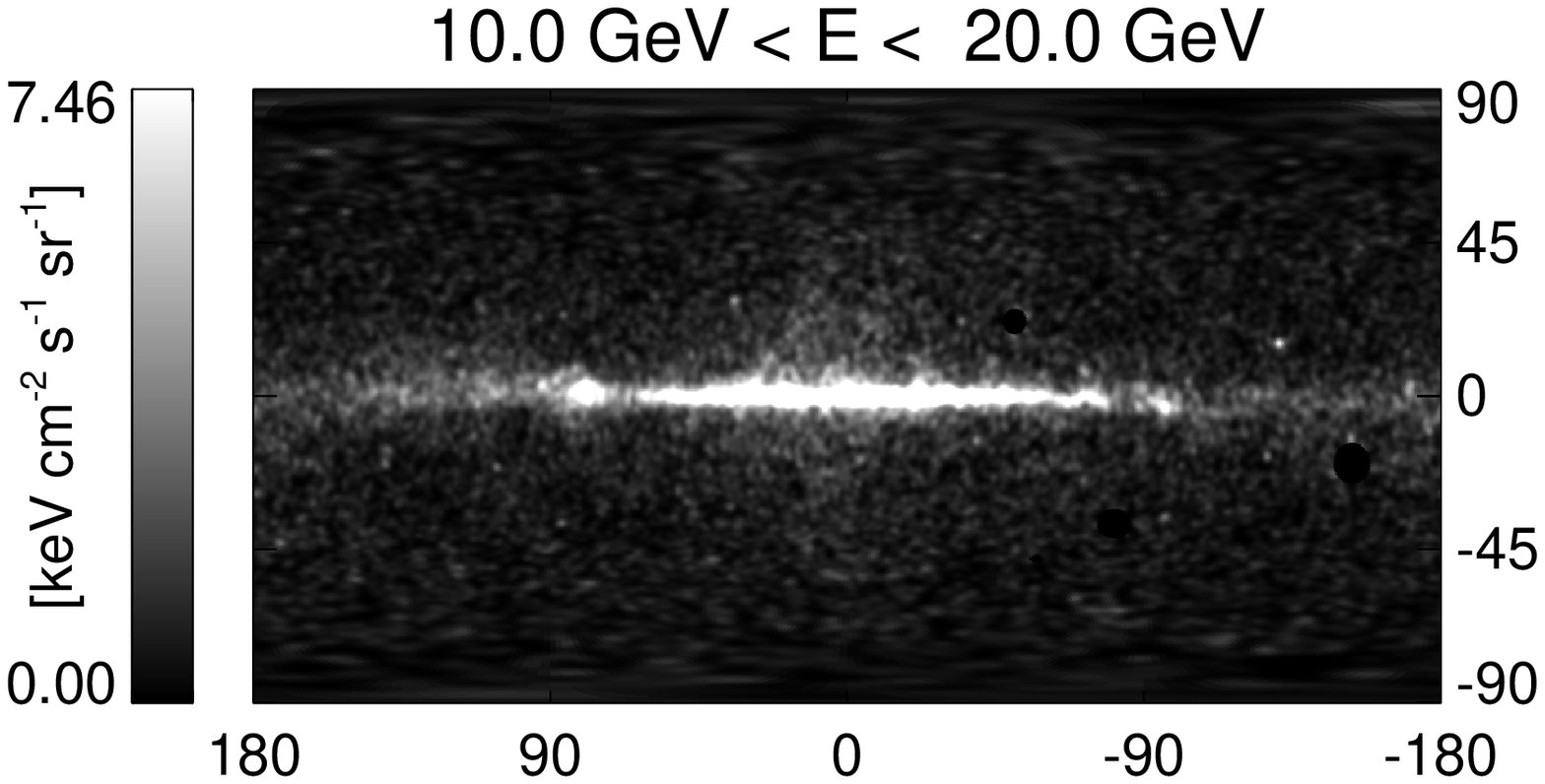}
  \includegraphics[width=0.49\textwidth]{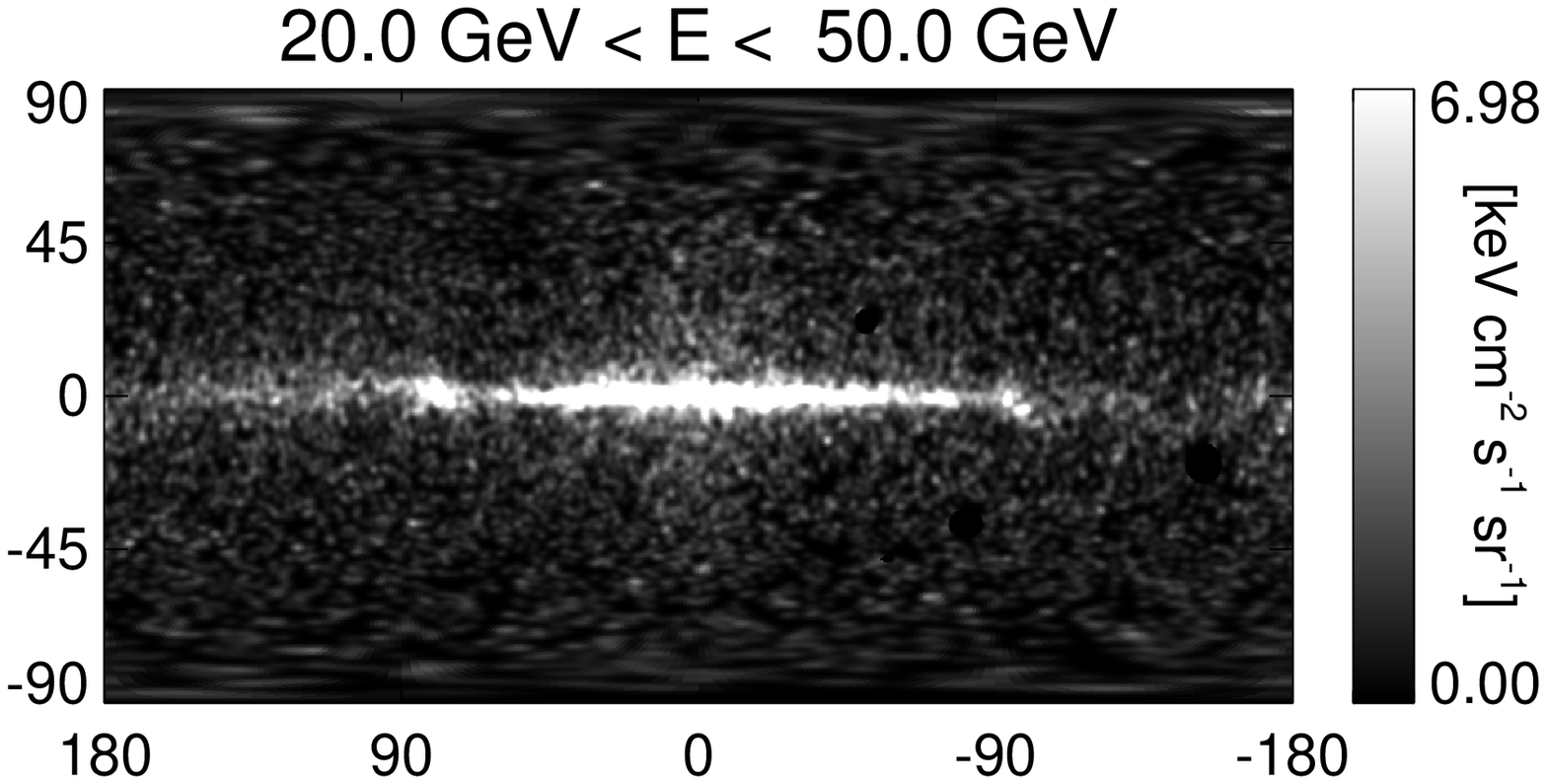}

  \includegraphics[width=0.49\textwidth]{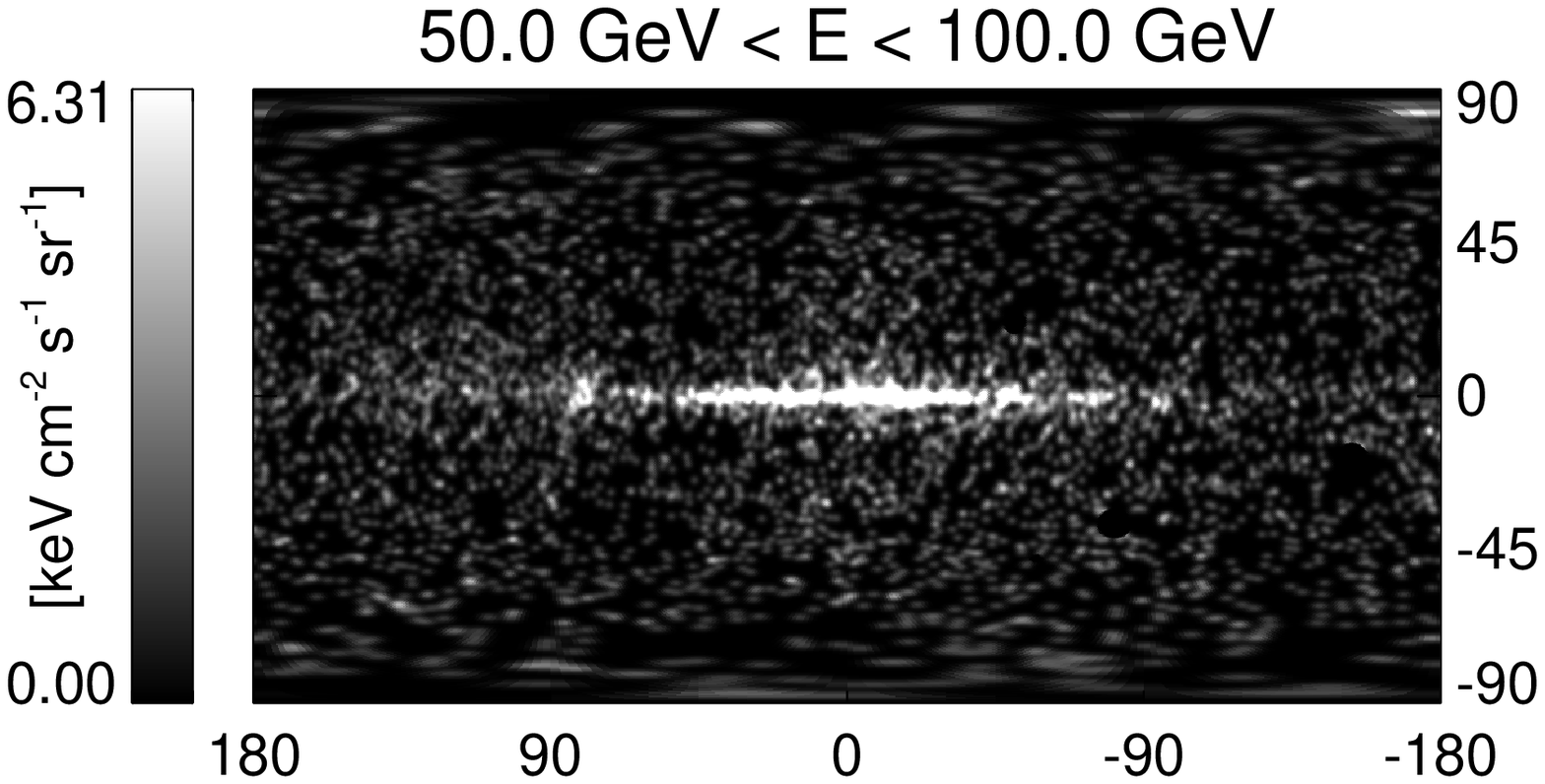}
  \includegraphics[width=0.49\textwidth]{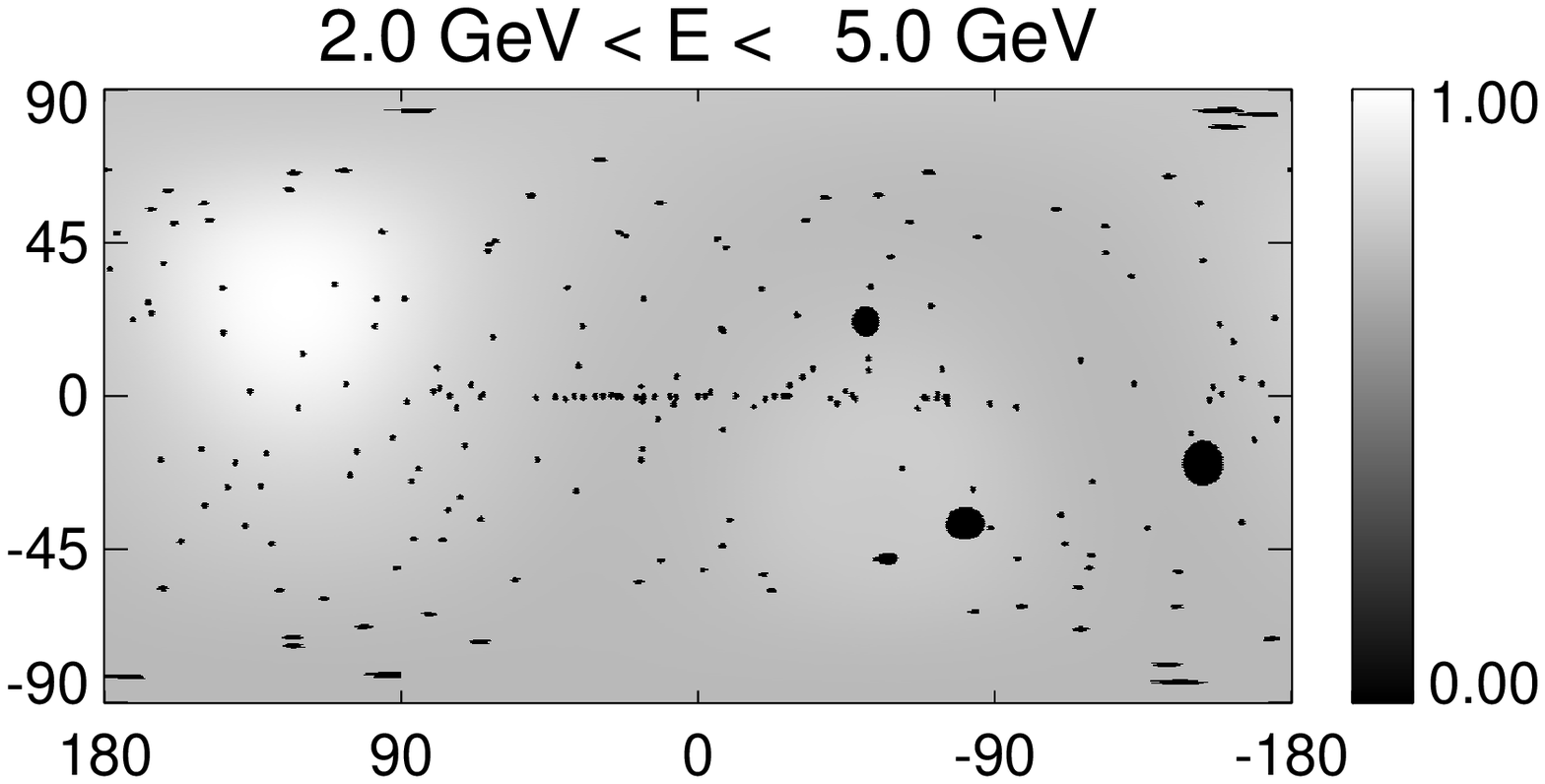}
\caption{ 
  \emph{Top and bottom left:} Full sky \emph{Fermi} $\gamma$-ray maps
  in various energy bins.  The mask includes the 3-month \emph{Fermi}
  point source catalog as well as the LMC, SMC, Orion-Barnard's
  Loop, and Cen A.  \emph{Bottom right:} Exposure time
  map with our mask overlaid and stretched to $0-100\%$ of the peak
  exposure time.  The variation in the exposure is a small modulation
  -- even setting it to unity does not change our qualitative results.
}
\label{fig:fermimaps}
\epm

Electron cosmic rays at $10-100$ GeV primarily lose energy in the
diffuse interstellar medium by producing synchrotron microwaves and
inverse-Compton (IC) scattered gammas.  Synchrotron losses are
proportional to magnetic field energy density, $U_B = B^2/8\pi$, while
IC losses are proportional to the interstellar radiation field energy
density, $U_{ISRF}$, in the Thomson limit, and less in the
Klein-Nishina limit.  Bremsstrahlung off the ambient gas also
occurs, but is expected to be sub-dominant in the regions of interest.
Therefore, the best test of the haze hypothesis is to search for IC
gammas in the \emph{Fermi} data, which was studied in the context of
dark matter signals by
\cite{Cholis:2008wq,Zhang:2008tb,Borriello:2009fa,Cirelli:2009vg,Regis:2009md,Belikov:2009cx,Meade:2009iu}.

Previous studies of high latitude gamma-ray emission have reported
measurements of an excess of emission above the Galactic plane, most
notably SAS-2 \citep{Fichtel:1975fi,Kniffen:1981kn}, COS-B
\citep{Strong:1984cb}, and EGRET
\citep{Smialkowski:1997sm,Dixon:1998di}.  However, in each case the
experiments either covered an insufficient energy range (SAS-2 and
COS-B) or did not have sufficient sensitivity and angular resolution
(EGRET) to permit a spatial correlation with the WMAP haze.
\emph{Fermi} overcomes both of these obstacles and allows us to search
for the gamma-ray counterpart to the microwave haze for the first
time.

The presence of an IC signal at the expected level can confirm that
the microwave haze is indeed synchrotron, ruling out the first null
hypothesis.  From the spectrum of the IC, we can estimate the
electron spectrum required to make the signal, addressing the second
null hypothesis.  For example, the presence of $\sim 50~\GeV$ IC
photons requires electrons of $E>50~\GeV$, perhaps much greater.
Furthermore, IC photons provide valuable information about the
spatial distribution (disk vs.\ bulge) of the source of these
particles, which in turn can constrain hypotheses about their
origin.\footnote{The
  synchrotron haze depends on the Galactic magnetic field while the
  IC haze depends on the Galactic ISRF, and so the two morphologies
  should not be identical, and could be quite different.}

In \S 2 we briefly review the \emph{Fermi} data, describe our
map-making procedure, and display full-sky maps at various energy
ranges.  In \S 3 these maps are analyzed with template correlation
techniques and resultant residual maps and spectra are shown.
Finally, \S 4 presents our interpretation of the signals and discusses
potential sources of contamination.  Estimates for the possible contamination from unresolved point sources are given in \refapp{pointsources}. Appendix \ref{sec:datarelease} details the creation and processing of the gamma-ray sky maps used in this analysis, and provides instructions for downloading them. Appendix \ref{sec:likelihood} contains a discussion of Poisson likelihood analysis on smoothed maps.

\section{Data}
\label{sec:data}

The LAT (Large Area Telescope) on \emph{Fermi} (see
\citealt{Gehrels:1999ri} as well as the \emph{Fermi}
homepage\footnote{\texttt{http://fermi.gsfc.nasa.gov/}}) is a
pair-conversion telescope consisting of 16 layers of tungsten on top
of a calorimeter with a thickness of 7 radiation lengths.  The entire
instrument is wrapped in a scintillating anti-coincidence detector to
provide a particle veto.  The spacecraft occupies a low Earth orbit
with an inclination of $25.6\degree$.  The field of view is so wide
that the entire sky may be covered in two orbits by rocking the
spacecraft north of zenith on one orbit and south of zenith on the
other.  Several times per month, \emph{Fermi} interrupts this pattern
to point the LAT at a gamma-ray burst, though this has little impact
on the integrated exposure map.  When the spacecraft passes through
the South Atlantic Anomaly (SAA), CR contamination increases and
significant data must be discarded, reducing the mean exposure at
southern declinations.  Beyond a zenith angle of $105\degree$ the data
are significantly contaminated by atmospheric gammas.  We excise such
data, and select only events designated ``Class 3'' (diffuse class) by
the LAT pipeline.  The LAT collaboration plans to release a cleaner
class of events in the future, however, at the time of this writing,
the Class 3 events are the most likely to be real diffuse gamma-ray
events.

The events are then binned into a full sky map using HEALPix, a
convenient iso-latitude equal-area full-sky pixelization widely used
in the CMB community.\footnote{HEALPix software and documentation can be found at
  \texttt{http://healpix.jpl.nasa.gov}, and the IDL routines used in
  this analysis are available as part of the IDLUTILS product at
  \texttt{http://sdss3data.lbl.gov/software/idlutils}.} Spherical
harmonic smoothing is straightforward in this pixelization, and we
smooth the maps to a Gaussian PSF, usually of $2\degree$ FWHM.  The
full-sky \emph{Fermi} maps are displayed in \reffig{fermimaps} along
with an exposure map.  See Appendix \ref{sec:datarelease} for more
details on map construction, smoothing, masking, and for instructions
on how to download the maps.

\section{Analysis}
\label{sec:analysis}

In this work our goal is to test our general preconceptions about
what gamma-ray signals should be present, and identify any unexpected features in the \emph{Fermi} data; we avoid detailed comparisons between the data and specific theoretical models for the Galactic gamma-ray emission. Our approach is to compute linear
combinations of \emph{Fermi} maps at several energies and perform
template analyses with maps of the ISM, radio maps, etc.\ to see what
emerges.  This sort of open-minded analysis is flexible enough to find
the unexpected.   

An alternate approach would be to attempt to fit the data with a sophisticated physical model, in the context of some simulation code
(e.g. GALPROP).  Such a physical model can be quite detailed,
including the 3D distributions of gas and dust, the 3D distribution of
optical and FIR photons density and direction, the 3D magnetic field,
and a 3D model of p, e$^-$ injections.  By propagating these primary
particles with GALPROP, the resulting $\pi^0$, bremsstrahlung and IC
signals may be predicted and compared with \emph{Fermi}. However, while a detailed physical model will certainly be crucial to a full understanding of the \emph{Fermi} data (and such modeling is currently underway within the \emph{Fermi} collaboration), this approach may lack the flexibility to identify new emission components that cannot be absorbed by modifying parameters in the model.
On the other hand, such a model may also have too much freedom, so that meaningful patterns
are absorbed into the fit and left unnoticed.  In future work, we will take the signals revealed by
our initial analysis and fold them back into a full physical model.

\subsection{Diffuse Gamma Templates}

There are three well-known mechanisms for generating gamma-rays at the
energies observed by \emph{Fermi}.  First, at low ($\sim$1 GeV)
energies, gamma-ray emission is dominated by photons produced by the
decay of $\pi^0$ particles generated in the collisions of cosmic ray
protons (which have been accelerated by SNe) with gas and dust in the
ISM.  Second, relativistic electrons colliding with nuclei (mostly
protons) in the ISM produce bremsstrahlung radiation.  Finally, those
same electrons interact with the interstellar radiation field (ISRF)
and inverse Compton scatter CMB, infrared, and optical photons
up to gamma-ray energies.  A schematic of the relative importance of
these emission mechanisms in the Galactic plane ($|\ell| \leq 30$,
$|b| \leq 5$) generated by the GALPROP code, version 50p
\citep{Strong:1998fr,Porter:2005qx,Strong:2007nh} is shown in
\reffig{galprop_example}.

\bp
\centerline{
  \includegraphics[width=0.49\textwidth]{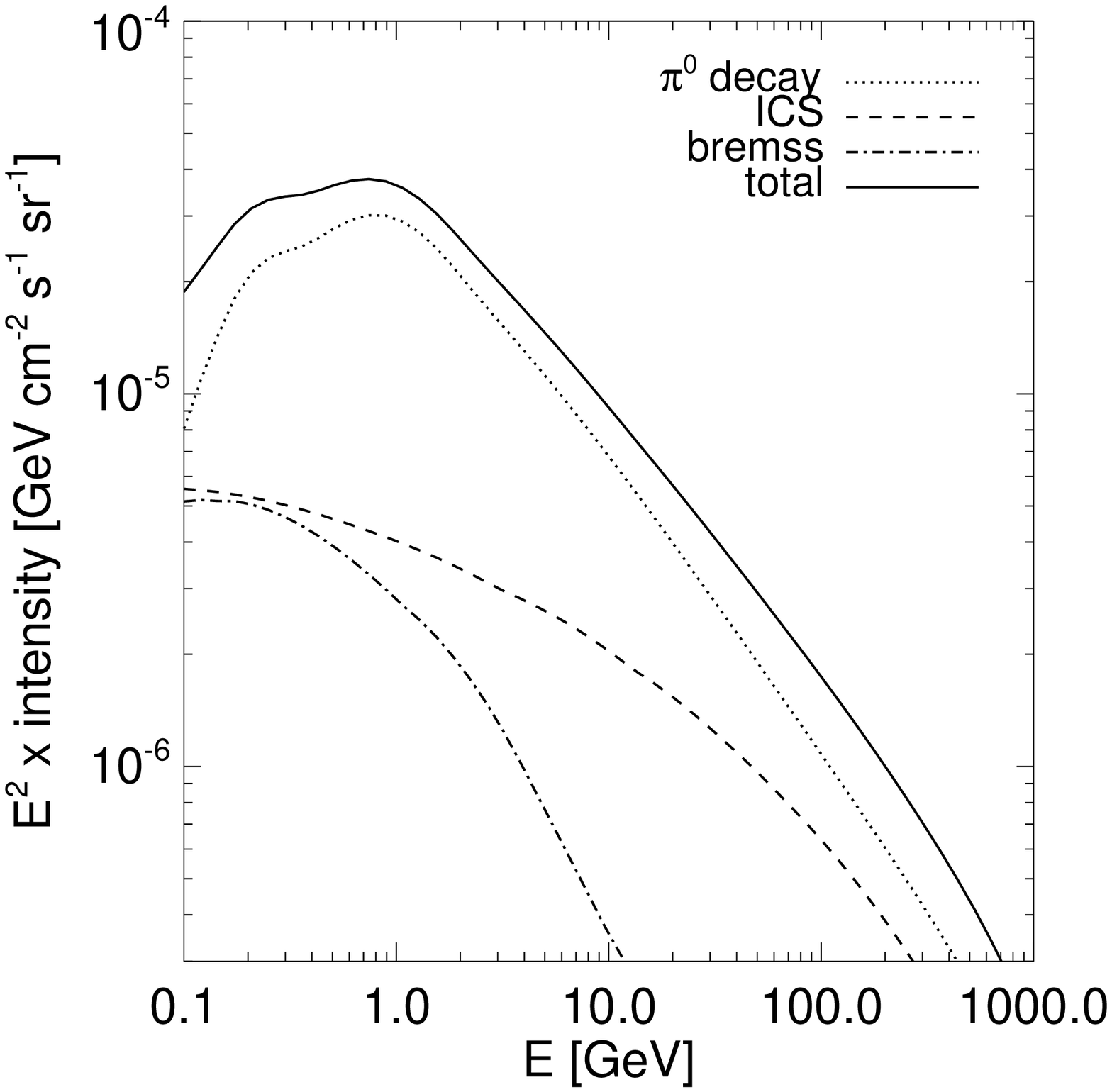}
}
\caption{ 
  GALPROP model illustrating the three primary gamma-ray emission
  mechanisms (see \refsec{analysis}) and their relative amplitudes in
  the Galactic plane ($|\ell| \leq 30$, $|b| \leq 5$).
}
\label{fig:galprop_example}
\ep

\bpm
\centerline{
  \includegraphics[width=0.9\textwidth]{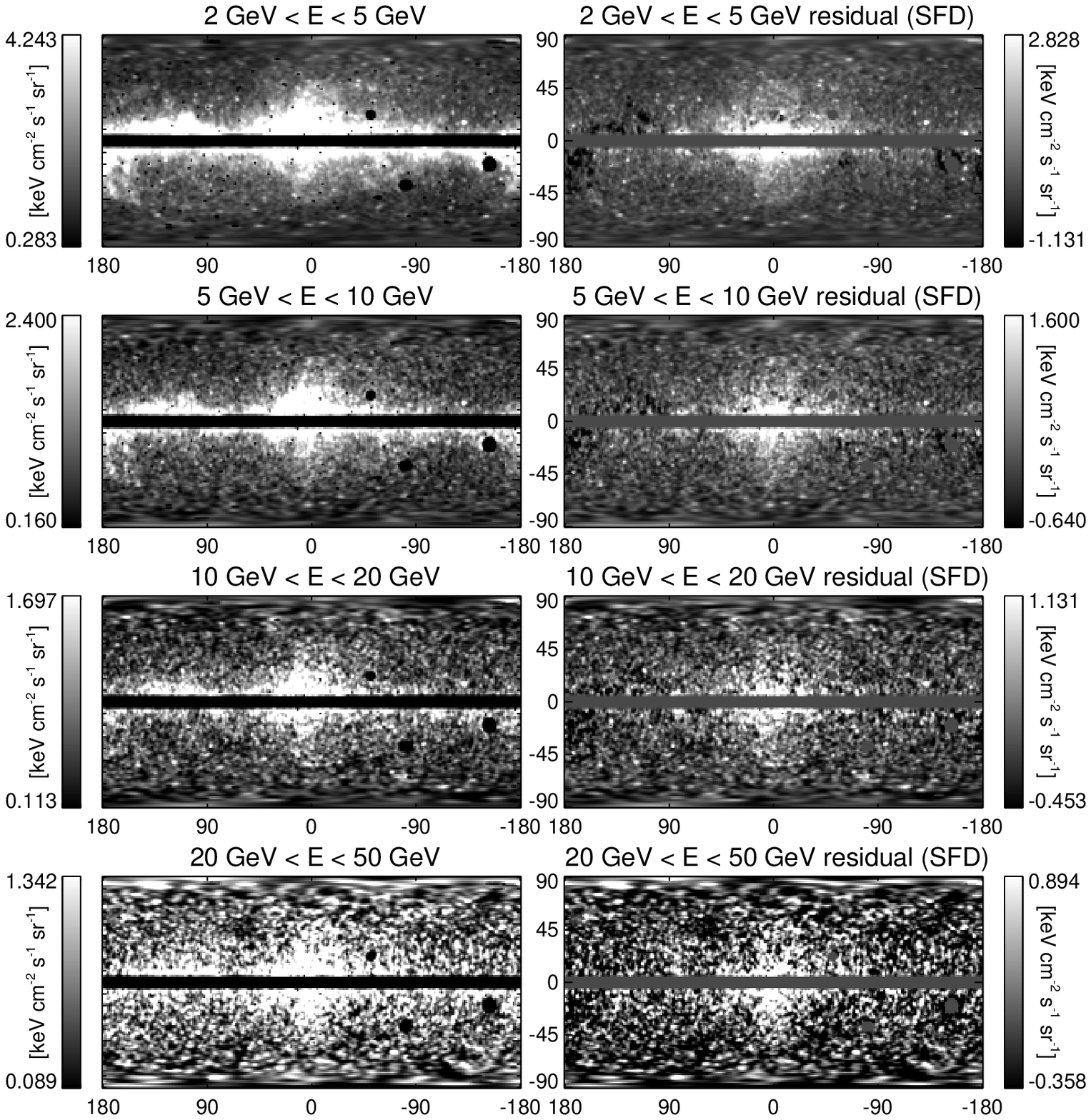}
}
\caption{ 
  Residual maps after subtracting the SFD dust template from 
  \emph{Fermi} maps at various energies.  The mask is described in
  Appendix \ref{sec:datarelease}.  Cross-correlations are done over unmasked
  pixels and for $75 \leq \ell \leq 285$.  Although the template
  removes much of the emission, there is a clear excess towards the
  Galactic center.  This excess also includes a disk-like component which
  is likely due to IC and bremsstrahlung from softer electrons (see
  \reffig{fermi_sub}).
}
\label{fig:fermi_sub_sfd}
\epm

\bpm
\centerline{
  \includegraphics[width=0.49\textwidth]{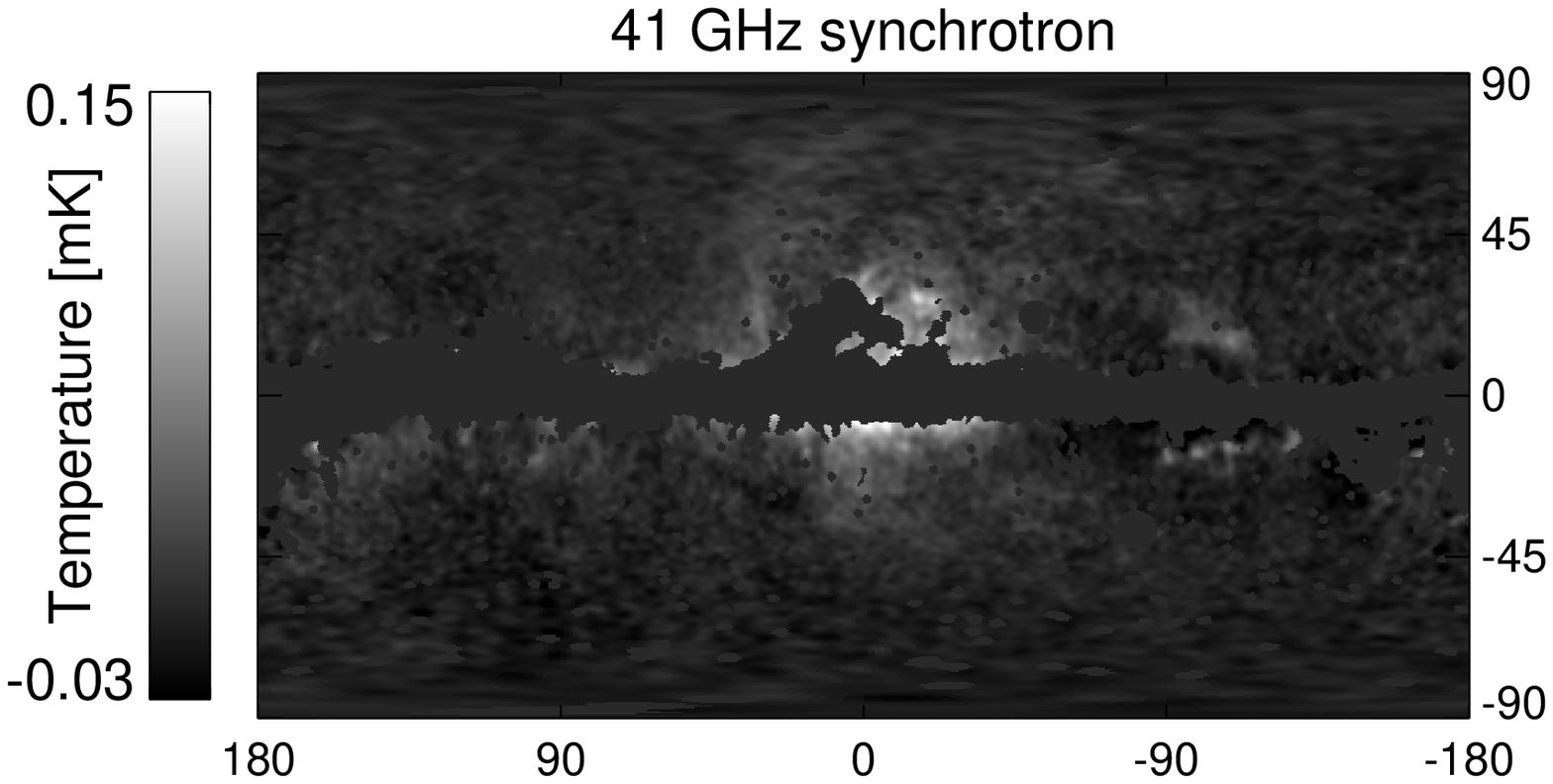}
  \includegraphics[width=0.49\textwidth]{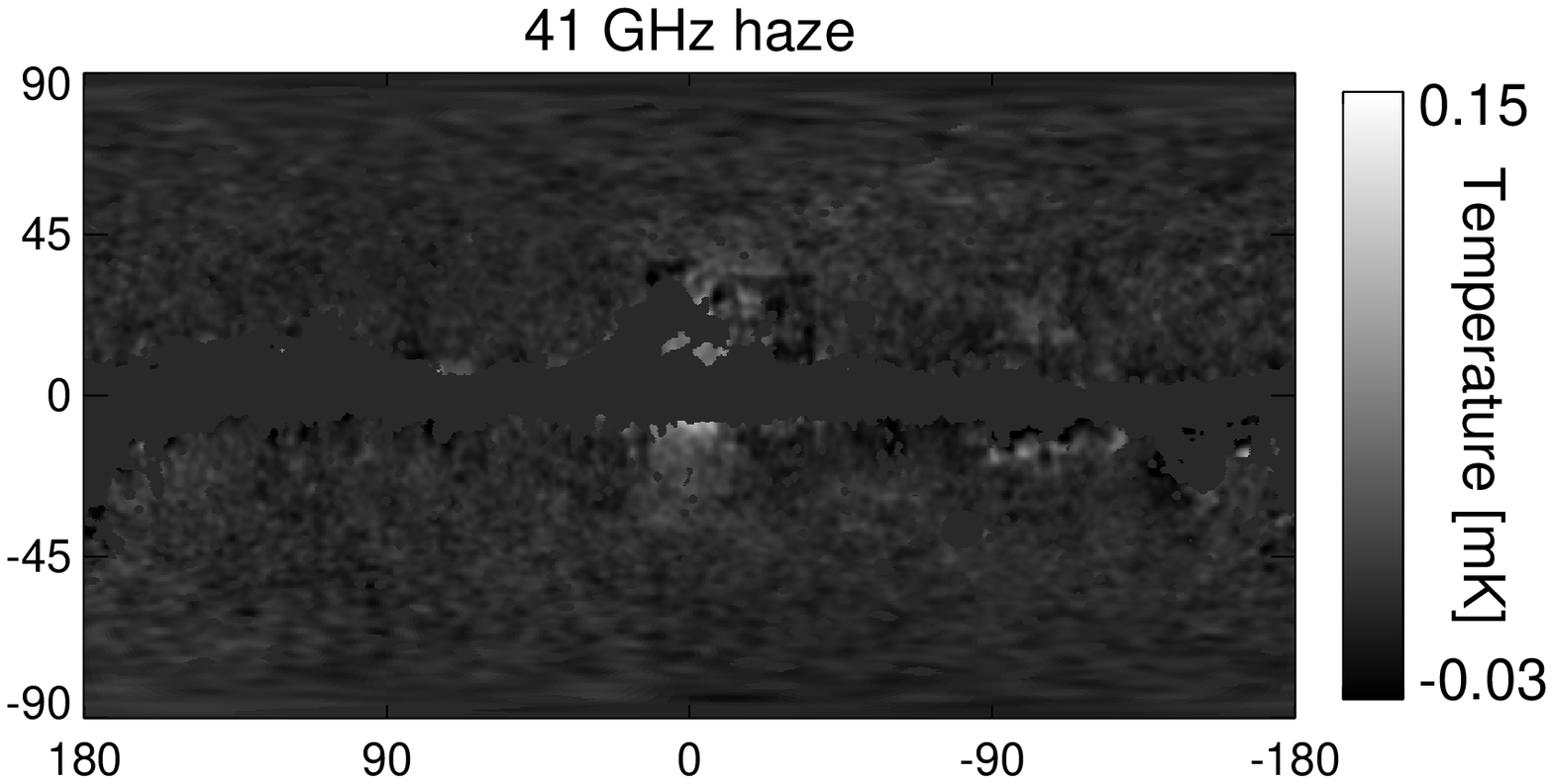}
}
\centerline{
  \includegraphics[width=0.49\textwidth]{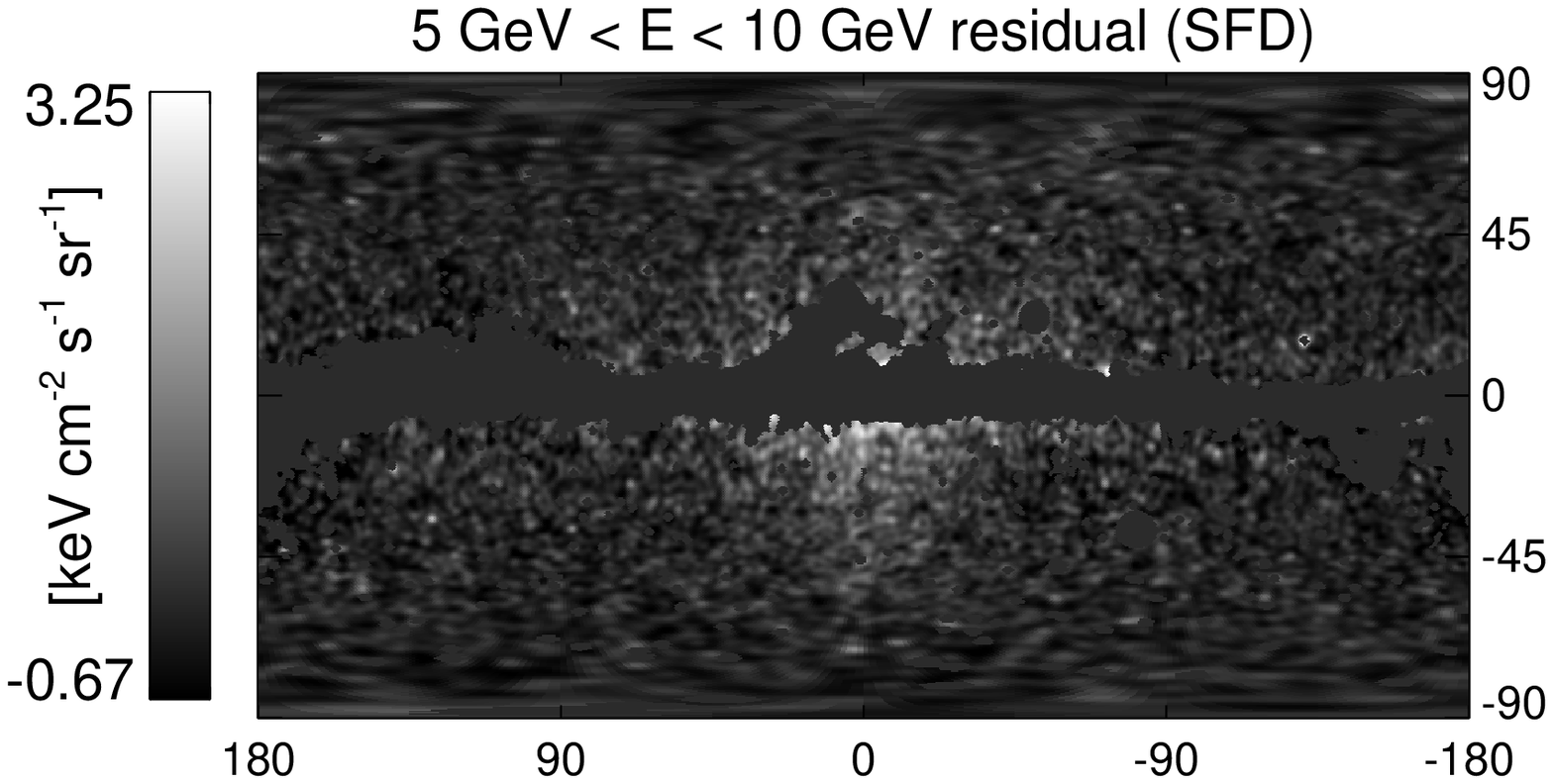}
  \includegraphics[width=0.49\textwidth]{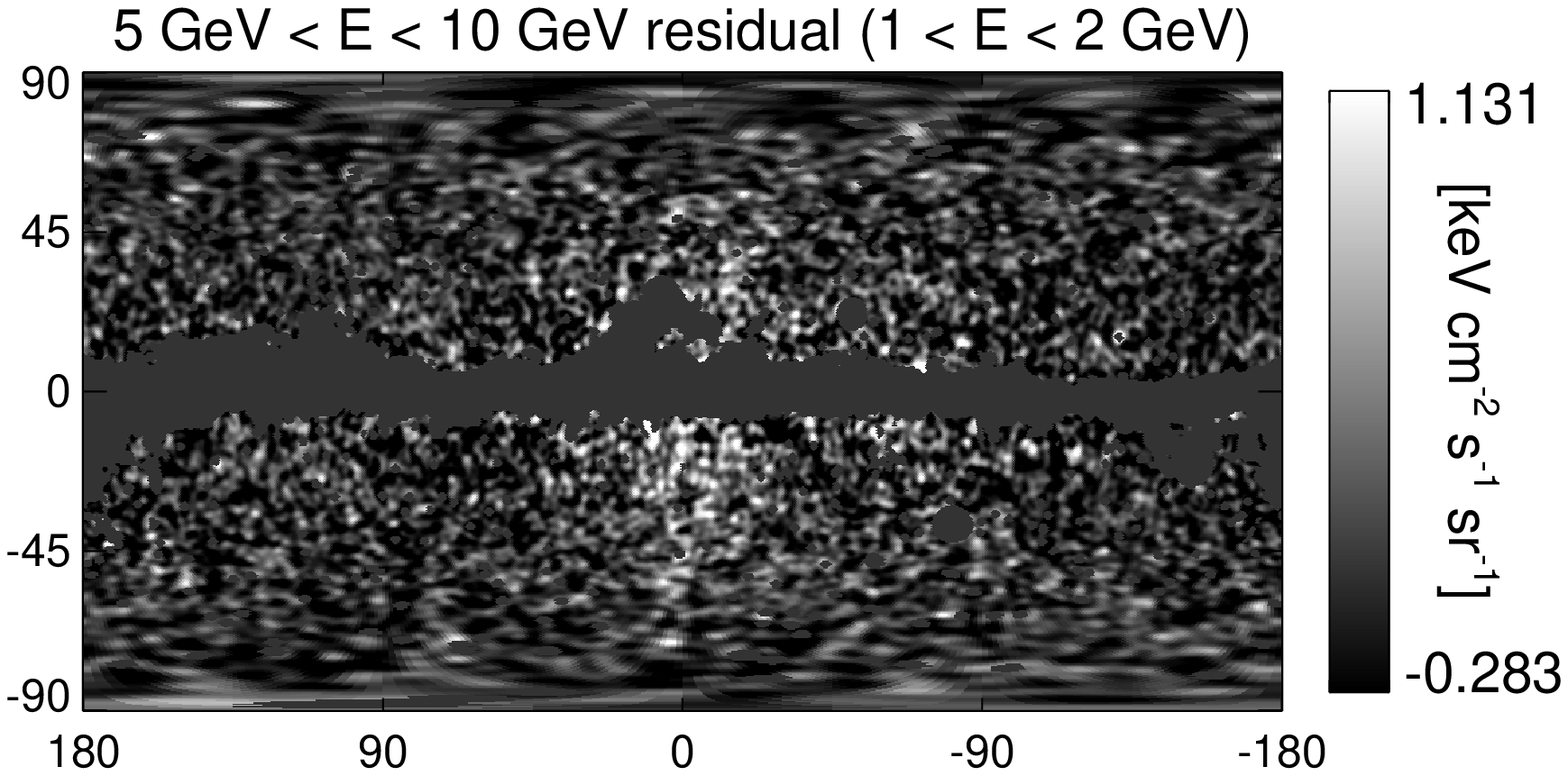}
}
\caption{ 
  \emph{Top:} the WMAP Q-band (41 GHz) total synchrotron (left) and
  haze (right).  \emph{Bottom:} the residual map obtained with the
  Type 1 template fit using the SFD map (left) and the 1-2 GeV
  \emph{Fermi} map (right) as a template for $\pi^0$ decay emission.
  These residual maps are the same as the right column second row
  panel of \reffig{fermi_sub_sfd} and \reffig{fermi_sub} respectively
  though with a different stretch.  Both maps are shown with the mask
  used in the \cite{Dobler:2008ww} microwave analysis for comparison.
  The upper panel represents cosmic ray electrons interacting with the
  Galactic magnetic field to produce synchrotron while the lower panel
  represents cosmic ray electrons interacting with the ISRF to produce
  IC emission, and colliding with the ISM to produce bremsstrahlung.  The morphological similarities between the
  microwave haze and gamma-ray haze (right) are striking, indicating a
  correspondence between the two electron populations.
}
\label{fig:fermi_comp_wmap}
\epm

Since $\pi^0$ gammas and bremsstrahlung are produced by interactions of
protons and electrons (respectively) with the ISM, these emission
mechanisms should be morphologically correlated with other tracers of
the ISM, such as the SFD dust map based on 100$\mu$ far IR data \citep{Schlegel:1997yv}.
The $\pi^0$ gamma-ray intensity scales with the ISM volume density
times the proton CR density, integrated along the line of sight.  In
the limit where the proton cosmic ray spectrum and density is
spatially uniform, the ISM column density is a good tracer of $\pi^0$
emission.  Likewise, for a uniform electron spectrum, it is a good
tracer of bremsstrahlung.  Because our analysis is limited to $|b|
>5\degree$, much of the emission we see is within a few kpc, so the
assumption of uniform CR density is more valid than it would be for
the entire Galaxy, particularly for protons, which have much larger propagation
lengths than electrons.

\subsection{Residual Maps}
\label{sec:residualmaps}

Since our goal is to search for an IC emission component with a
morphology which roughly matches the microwave haze (i.e., centered on
the GC, \emph{roughly} spherical, and about 20-40$\degree$ in radius),
we now attempt to remove the $\pi^0$ emission from the maps shown in
\reffig{fermimaps}, using the same template fitting technique used in
\cite{Finkbeiner:2003im} and \cite{Dobler:2008ww}.  We perform
multiple types of template fits.  Type 1 uses only the \emph{Fermi}
map itself at 1-2 GeV which roughly traces $\pi^0$ emission
because the gamma ray sky at those energies is dominated by $\pi^0$
gammas (with subdominant contributions from bremsstrahlung and IC).  Type 2 uses only the SFD
dust map which is a direct tracer of ISM density and so should roughly
map where the $\pi^0$ gammas and much of the bremsstrahlung are produced -- again, up to some
uncertainty involving the line of sight distribution.  In each case, a
uniform background is included in the fit, making our results
insensitive to zero point offsets in the maps.

We model the \emph{Fermi} map at energy $E$, $F(E)$, as a linear
combination of template maps, $F_{\rm model} = T c_T $, where $F_{\rm
  model}$ is a column vector of $N_{\rm pix}$ unmasked pixel values,
$T$ is the $N_{\rm template} \times N_{\rm pix}$ template matrix, and
the correlation coefficients $c_T$ are chosen to minimize the mean
squared residual,
\be
\langle (F-F_{\rm model})^2 \rangle = \langle (F-T c_T)^2 \rangle,
\ee
averaging over pixels. 
The least-squares solution,
\be
  c_{T}(E) = \left( T^{\rm T}T \right) ^{-1} \times \left( T^{\rm
    T}F(E) \right),
\label{eq:tempfit}
\ee
yields the template correlation coefficients at each energy.
In this fit, we mask out 
the \emph{Fermi} 3-month point source
catalog as well as the LMC, SMC, Orion-Barnard's Loop, and Cen A
(see Appendix \ref{sec:datarelease} for details). 
We also mask all pixels with Galactic latitude $|b|
< 5\degree$.  Cross-correlations are done over unmasked pixels and for
several different ranges in $\ell$: eight longitudinal slices that
have $|\ell| \geq 75\degree$ (to avoid the GC) and width $30\degree$
(Regions 1-7) as well as one region with $75\degree < \ell <
285\degree$ (Region 8).  Note that Region 8 is the union of Regions 1-7; 
Regions 1-7 are fit individually to show the variation in the fit spectrum, 
and Region 8 is fit to obtain the mean outer Galaxy signal. 
With these correlation coefficients, we
define the residual map to be,
\be
  R_{T}(E) = F(E) - c_{T}(E) \times T.
\ee
To the extent that the templates in $T$ match the morphology of the
$\pi^0$ and bremsstrahlung gammas, the residual map will include only
IC emission.

\reffig{fermi_sub_sfd} shows the resultant residual maps using the SFD
map as a morphological tracer of $\pi^0$ emission for the Region 8
fits.  The most striking feature of the difference maps is the
extended emission centered around the Galactic center and extending
roughly $40\degree$ in $b$.  The morphological correlation between the
WMAP synchrotron and the $R_{\rm SFD}({\rm 5-10\ GeV})$ is striking as
is shown in \reffig{fermi_comp_wmap}.  Here the 41 GHz synchrotron
\citep[haze plus Haslam-correlated emission,
  see][]{1982A&AS...47....1H, Dobler:2008ww} is shown side by side
with the 5-10 GeV \emph{Fermi} residual map with the mask used in the
\cite{Dobler:2008ww} microwave analysis overlaid for visualization.

\reffig{fermi_sub} shows residual maps using the 1-2 GeV \emph{Fermi}
maps as a template for the $\pi^0$ emission.  This sort of ``internal
linear combination'' has the advantage that $\pi^0$ emission cancels
out as long as the \emph{shape} of the proton CR spectrum is the same everywhere -- it
does not rely on the proton CR density to be uniform.  The residual
maps look largely similar to the case with the SFD template regressed
out, but there are some notable differences.  In particular, using
this 1-2 GeV template, the IC haze has a slightly ``taller''
appearance.  This seems to be due to a disk-like component that is present
in the 1-2 GeV maps but not in the dust map.
This is probably because the 1-2 GeV \emph{Fermi} map is
not entirely $\pi^0$ emission, but also contains bremsstrahlung and
IC components generated by electrons accelerated by SNe in the disk.  
The result is that when this lower energy map is regressed out
(i.e., cross-correlated and subtracted) from higher energy maps, this
emission component is subtracted along with the dust-correlated emission.
Conversely, the SFD map contains only emission from dust grains and
not relativistic electrons; while the bremsstrahlung from those electrons 
largely traces the gas distribution, the IC does not, and so is not regressed out
when using the SFD template.

\bpm
\centerline{
  \includegraphics[width=0.9\textwidth]{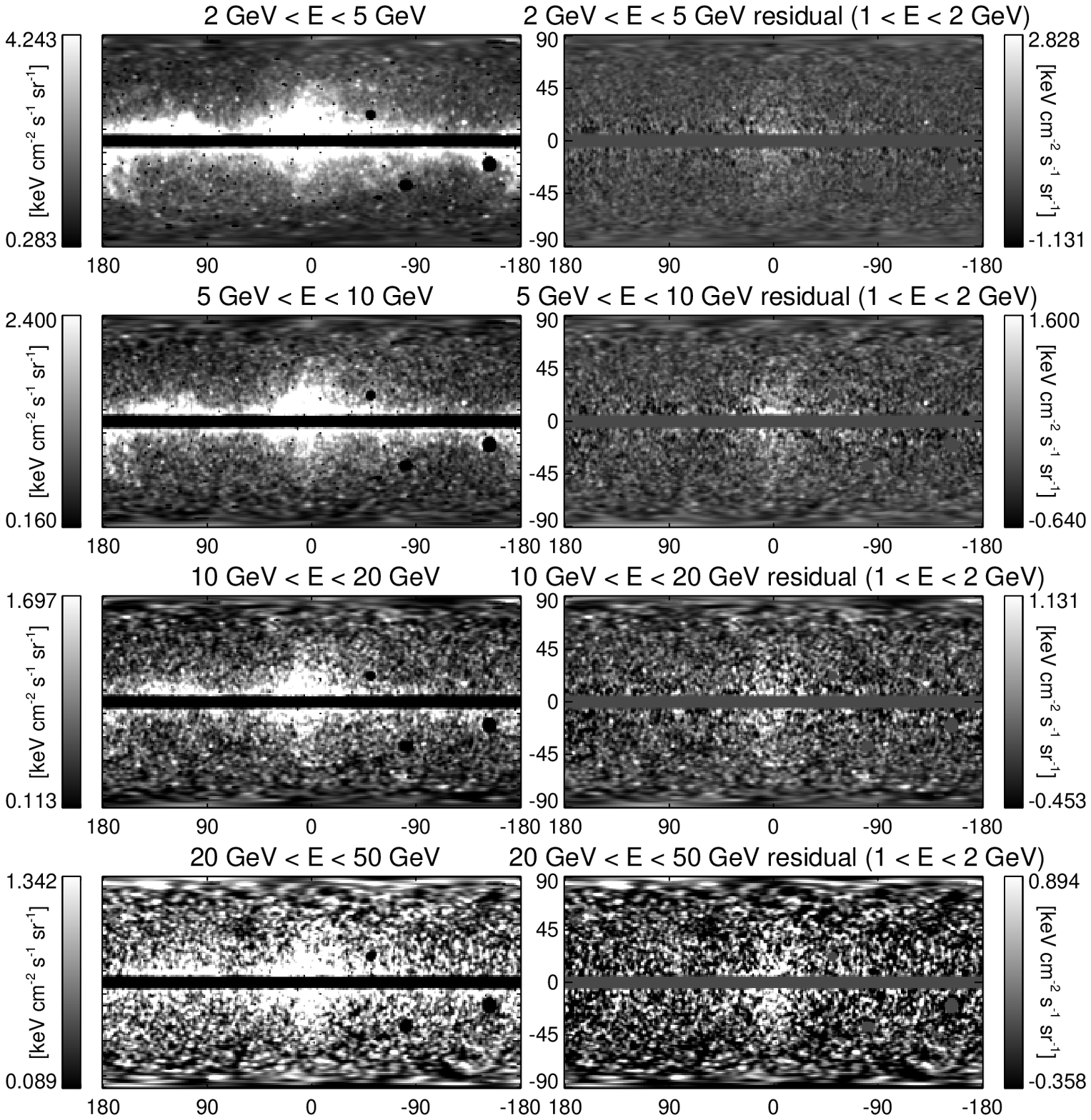}
}
\caption{ 
  The same as \reffig{fermi_sub_sfd} but using the \emph{Fermi} 1-2
  GeV map for cross-correlations instead.  Unlike the SFD dust map
  which should trace $\pi^0$ (and subdominant bremsstrahlung) emission only,
  the low energy \emph{Fermi} map includes the soft IC associated
  with lower energy electrons.  In fact comparing the residuals in
  this figure with those in \reffig{fermi_sub_sfd}, it is clear that
  the disk-like component has been subtracted leaving only the IC haze.
  Furthermore, the IC haze is more prominent in the high energy maps
  indicating a harder spectrum than $\pi^0$ emission which is the
  dominant emission mechanism at $\sim$1 GeV energies.
}
\label{fig:fermi_sub}
\epm

\bpm
\centerline{
  \includegraphics[width=0.49\textwidth]{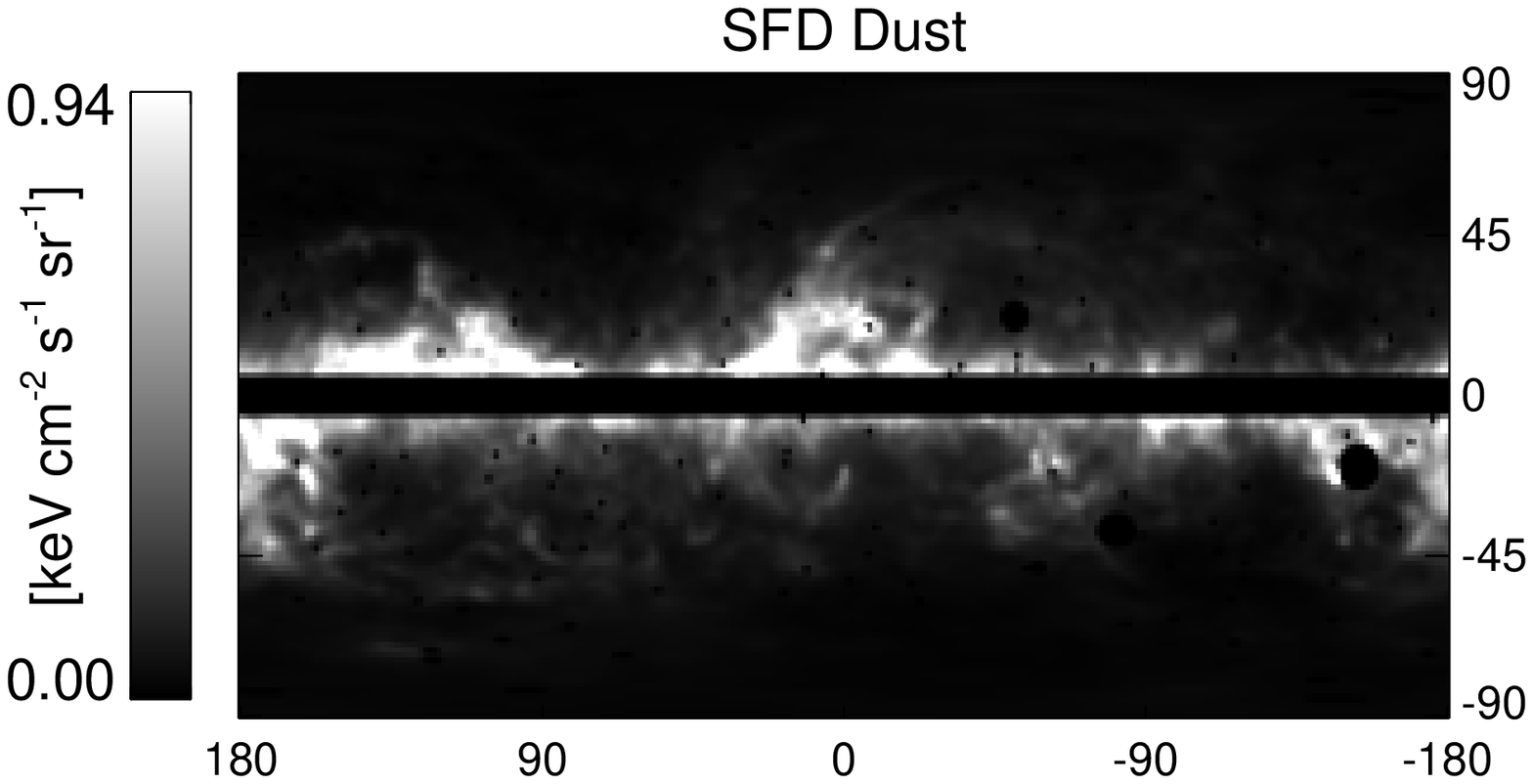}
  \includegraphics[width=0.49\textwidth]{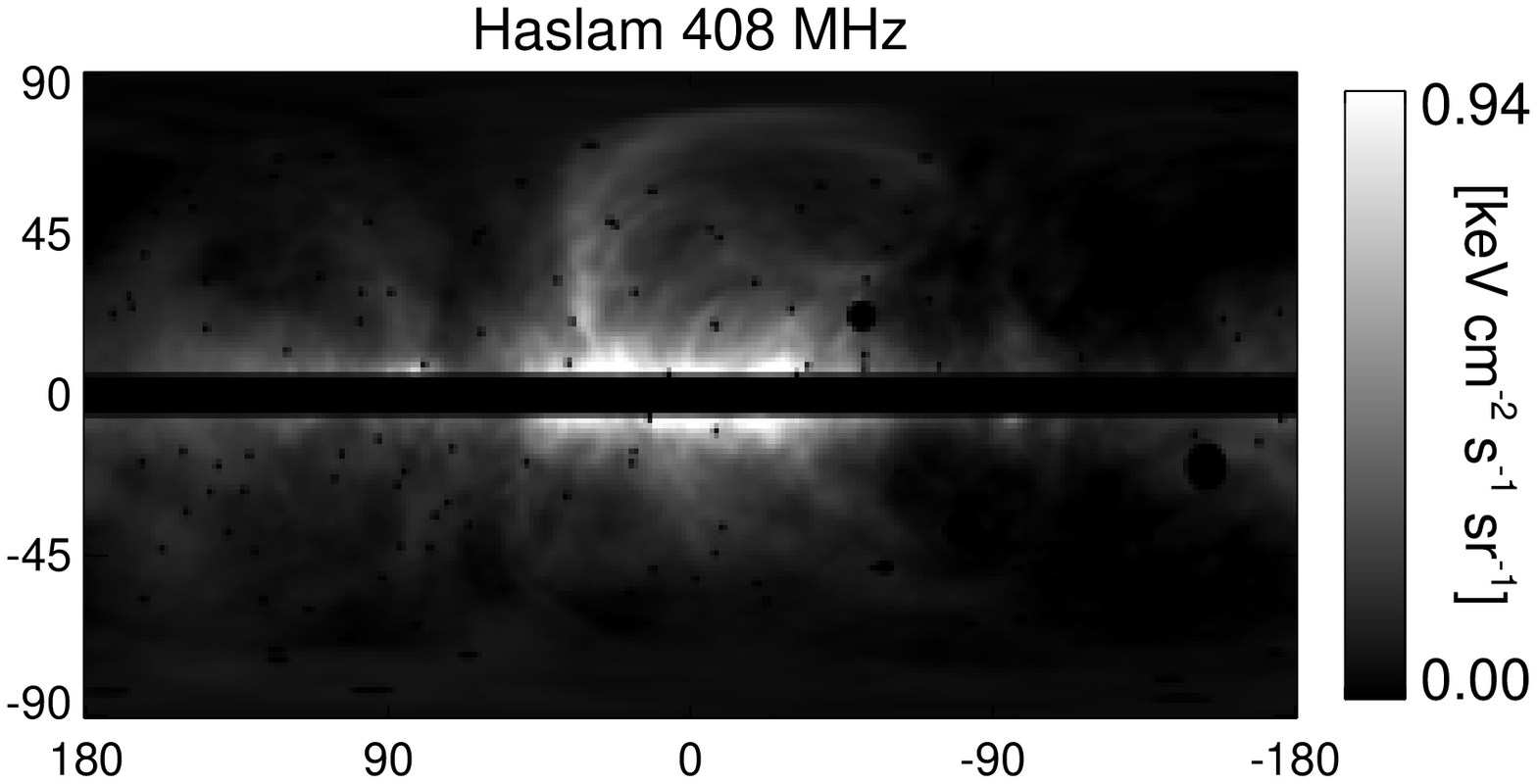}
}
\centerline{
  \includegraphics[width=0.49\textwidth]{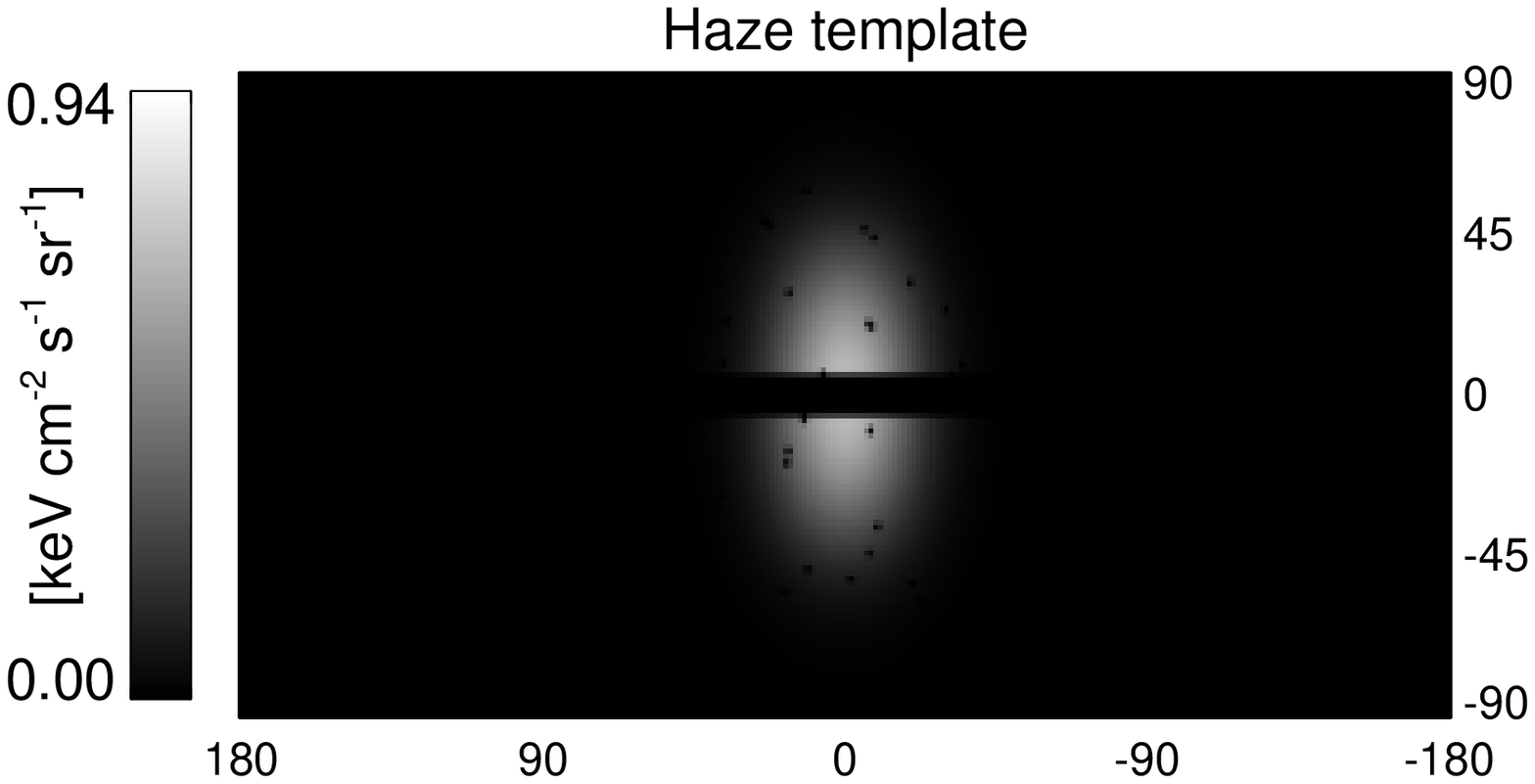}
  \includegraphics[width=0.49\textwidth]{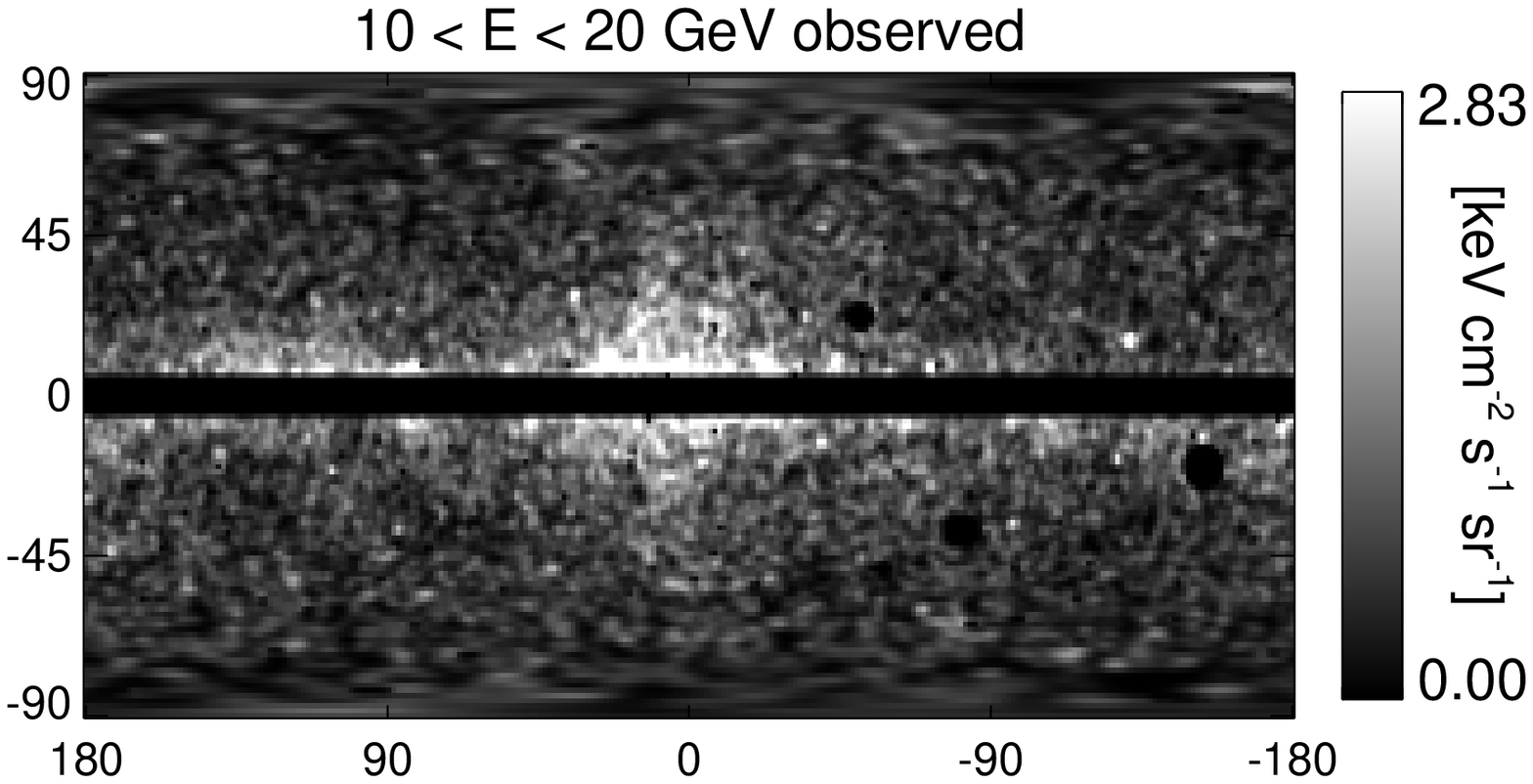}
}
\centerline{
  \includegraphics[width=0.49\textwidth]{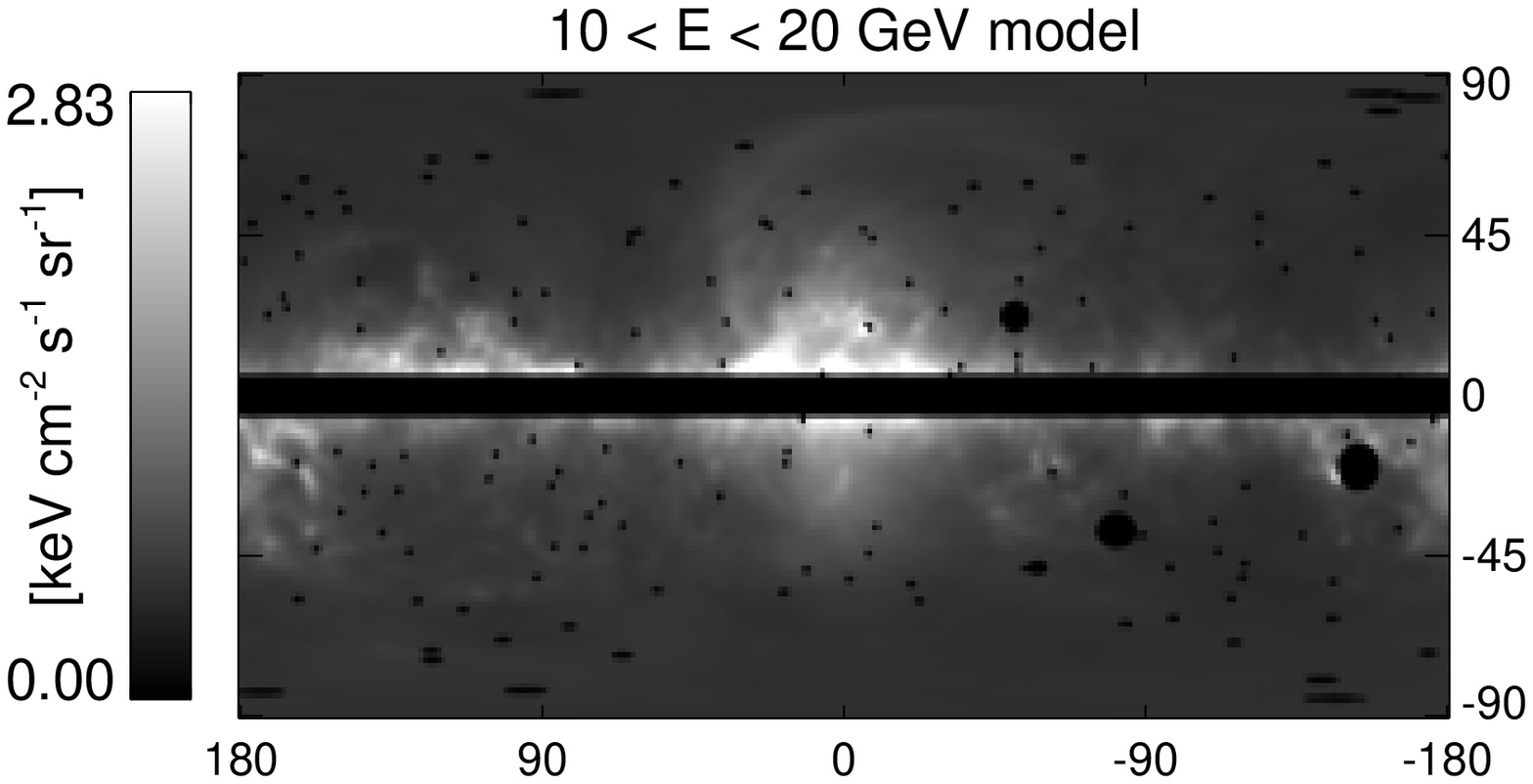}
  \includegraphics[width=0.49\textwidth]{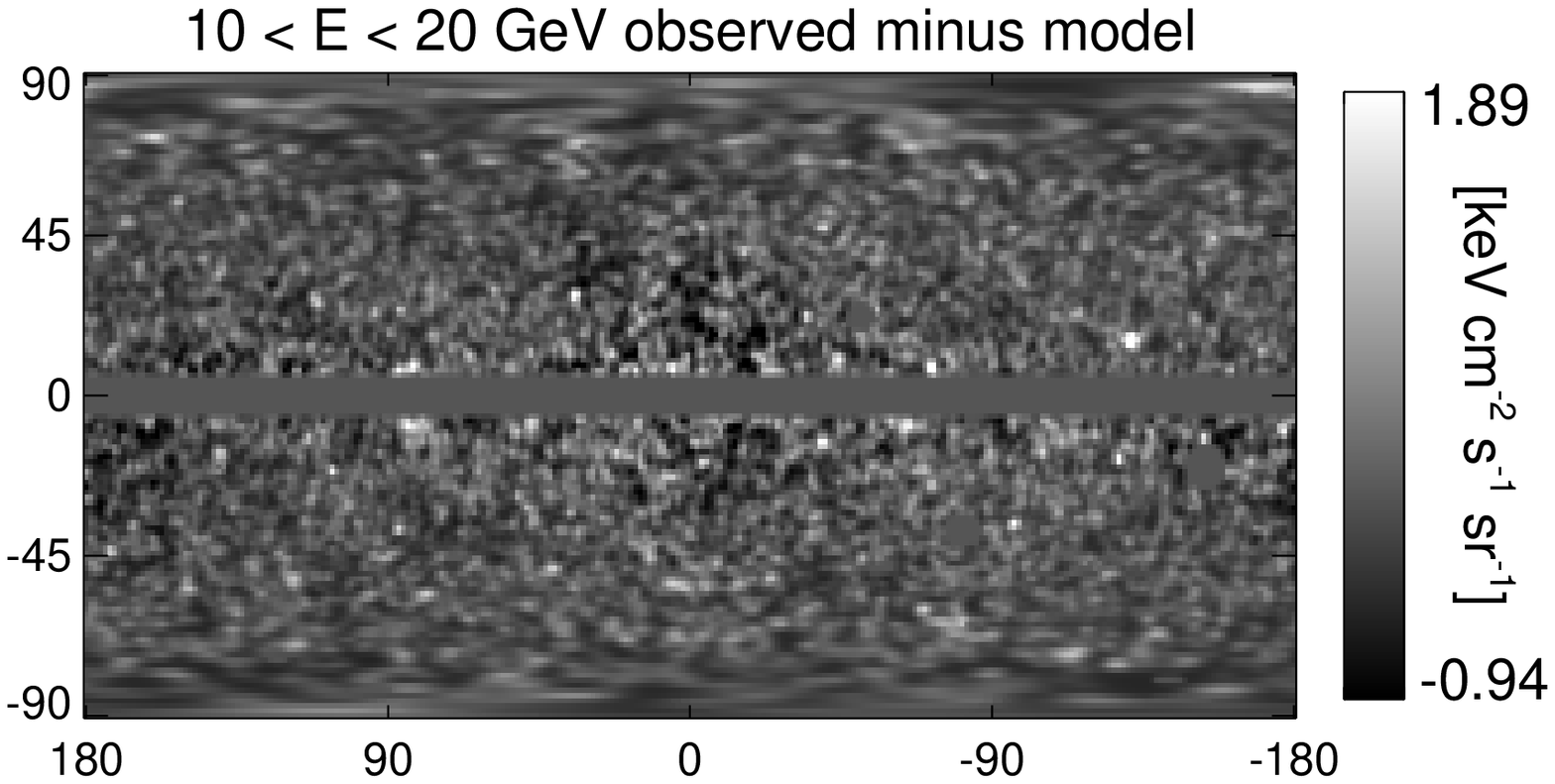}
}
\caption{ 
  The templates and fit solutions used in the Type 3 (see
  \refsec{residualmaps}) template fits.  \emph{Upper left:} the SFD
  dust map, \emph{upper right:} the Haslam 408 MHz map, \emph{middle
    left:} the bivariate Gaussian haze template, \emph{middle right:}
  the \emph{Fermi} map at 10-20 GeV (same as the first column, third
  row of \reffig{fermi_sub} but with a different stretch to show the
  detailed morphological structure), \emph{lower left:} the best fit
  template solution for the observed emission, \emph{lower right:} the
  residual map.  Note the very small residuals indicating that the
  template fit is a remarkably good representation of the data over
  large areas of the sky.
}
\label{fig:fermi_templates}
\epm

Despite the significant shot noise in the \emph{Fermi} map,
\reffig{fermi_comp_wmap} shows that there is a clear morphological
correlation between the microwave haze and the gamma-ray haze.  Of
course, we do not expect the morphologies to agree perfectly since the
microwave haze is generated by interactions of electrons with the
magnetic field while this \emph{IC haze} is due to interactions of
the electrons with the ISRF.  Nevertheless, this is evidence
that the microwave haze seen in the WMAP data is indeed
\emph{synchrotron} and is not some other component such as free-free
emission or spinning dust.

\subsection{Four-Component Template Fits}
\label{sec:fourcomponent}
Because the solution to Equation \ref{eq:tempfit} minimizes the
variance of the (zero mean) residual map $R(E)$, the presence 
of the IC haze affects the
coefficients $c_{\rm T}(E)$, since there is non-zero spatial
correlation between the templates and the IC haze.  To relax the
stress on the fit, we expand our template analysis with a third Type.
Type 3 uses four templates in $T$: the SFD map to trace $\pi^0$ and bremsstrahlung
emission, the 408 MHz \cite{1982A&AS...47....1H} map which is
dominated by radio wavelength synchrotron and thus roughly traces soft
spectrum electrons which produce soft IC and bremsstrahlung, and a
bivariate Gaussian of width $\sigma_\ell=15\degree$ and
$\sigma_b=25\degree$.  We note that this template is chosen to roughly
match the morphology in \reffig{fermi_sub} and has no other physical
motivation.  We also use a uniform template to fit out the
isotropic background signal in the maps, again, making our results
insensitive to zero points.  Lastly, for this fit we use \emph{all} values in $\ell$ (Region 9).

Note that since the bremsstrahlung originates from interactions of
the electrons with the ISM, its spatial distribution depends on both the gas density
and the cosmic ray electron density; consequently, some contribution from bremsstrahlung
will be present in both the SFD-correlated and Haslam-correlated emission.

The previous fits were done with uniform weighting and assuming
Gaussian errors, minimizing $\chi^2$.  For the Type 3 fit we do a more
careful regression, maximizing the Poisson likelihood of the 4-template model
in order to weight the Fermi data properly.  In
other words, for each set of model parameters, we compute the log
likelihood
\be
  \ln {\mathcal L} = \sum_i k_i\ln\mu_i - \mu_i - \ln(k_i!),
\ee
where $\mu_i$ is the synthetic map (i.e., linear combination of
templates) at pixel $i$, and $k$ is the map of observed counts.  Note
that the last term does not depend on the model parameters.  It may appear
strange at first to compute a Poisson likelihood on smoothed maps,
however, the smoothing is necessary to match PSFs at different
energies and with various templates (some of which have lower
resolution than \emph{Fermi} in the energy range of interest).  The smoothing itself does not pose any
problems for relative likelihoods, as we show in Appendix \ref{sec:likelihood}.

However, we must keep in mind that the uncertainties derived in this
way are the formal errors corresponding to $\Delta\ln {\mathcal L} = 1/2$, which
would be $1\sigma$ in the case of Gaussian errors.  The error bars
plotted are simply the square root of the diagonals of the covariance
matrix.  This estimate of the uncertainty should be accurate at high
energies, where photon Poisson noise dominates.  At low energies,
although the formal errors properly reflect the uncertainty in the fit
coefficients for this simple model, the true uncertainty is dominated
by the fact that the 4-template model is not an adequate
representation of the data.

\reffig{fermi_templates} shows the skymaps and best fit solution
including the residual map at 10-20 GeV while
\reffig{fermi_template_residual} shows residual maps at other
energies.  It is clear from these residuals that the template fitting
produces a relatively good approximation of the gamma-ray data over
large areas of sky.  Furthermore, \reffig{fermi_template_residual}
shows that \emph{not} including the bivariate Gaussian template for
the IC haze yields a statistically significant residual towards the
center indicating that a model including an IC haze is a better match
to the data then one without.  The prominent North Polar spur feature
in the Haslam map, which is thought to originate from synchrotron
emission from electrons in Loop I \citep{Large:1962}, is
over-subtracted in each case, because the North Polar spur is brighter
in the Haslam map than in the gamma-ray maps (i.e. the ratio of
synchrotron microwaves to IC gamma rays in the North Polar spur is
larger than in the rest of the Haslam map).\footnote{See
  \cite{Casandjian:2009} for a discussion of gamma-rays from Loop I
  seen by \emph{Fermi}.} This may be due to different ISRF, B-field and
ISM density values in Loop I relative to the inner Galaxy (since Loop
I is thought to be quite nearby), or may be due to a softer-than-usual
electron spectrum in Loop I, since the electrons producing the
synchrotron measured in the Haslam map are much lower energy than
those producing IC gamma-rays at energies measured by \Fermi.

\reffig{fermi_template_residual_smth} shows these same residual maps,
but with a smoothing of $10\degree$ which is on the order of the scale
of the haze emission.  With this large smoothing, smaller scale
variations (due to individual photons at high energies) are smoothed
over and the residual maps clearly show that the haze is a robust
feature at all energies.

In \reffig{fermi_template_residual_loE} we show the results of a four
template fit using the 1-2 GeV map instead of SFD to trace the $\pi^0$
emission.  This figure shows that the haze is \emph{not} due to the
SFD template being an imperfect tracer of $\pi^0$ emission.  If the
proton cosmic ray density is higher towards the GC, then SFD may
systematically underestimate the $\pi^0$ emission there and perhaps
the gamma-ray haze is the result.  However,
\reffig{fermi_template_residual_loE} shows that this is clearly not
the case.  The 1-2 GeV template includes the effects of proton cosmic
ray density variations (as well as line of sight gas density effects)
and the haze remains as a robust residual.  That is, the haze is
\emph{not} due to imperfect templates as suggested by
\cite{Linden:2010ea}.

\bpm
\centerline{
  \includegraphics[width=0.9\textwidth]{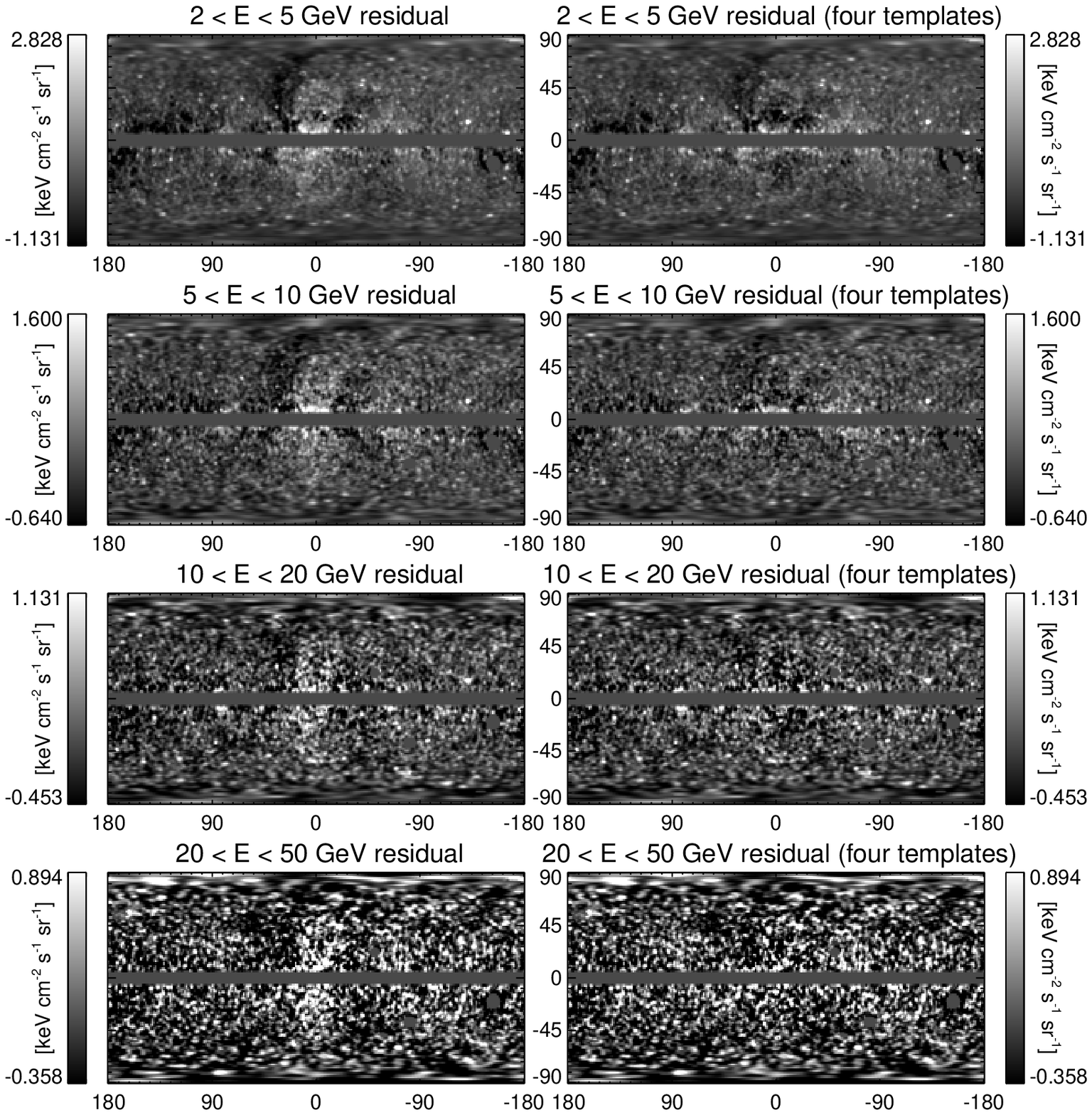}
}
\caption{ 
  Residual maps using the Type 3 template fit.  The right column is
  the same as the lower right map in \reffig{fermi_templates} but for
  maps at different energy bands.  The left column performs the same
  fit \emph{without} including a bivariate Gaussian template for the
  IC haze.  It is clear that not including the haze template results
  in a significant residual towards the GC in each energy band, but
  particularly at high energies.  Including the haze template improves
  $\ln{\mathcal L}$ by 504, 215, 78, and 54, respectively, for the
  4 energy bins shown.
}
\label{fig:fermi_template_residual}
\epm

\bpm
\centerline{
  \includegraphics[width=0.9\textwidth]{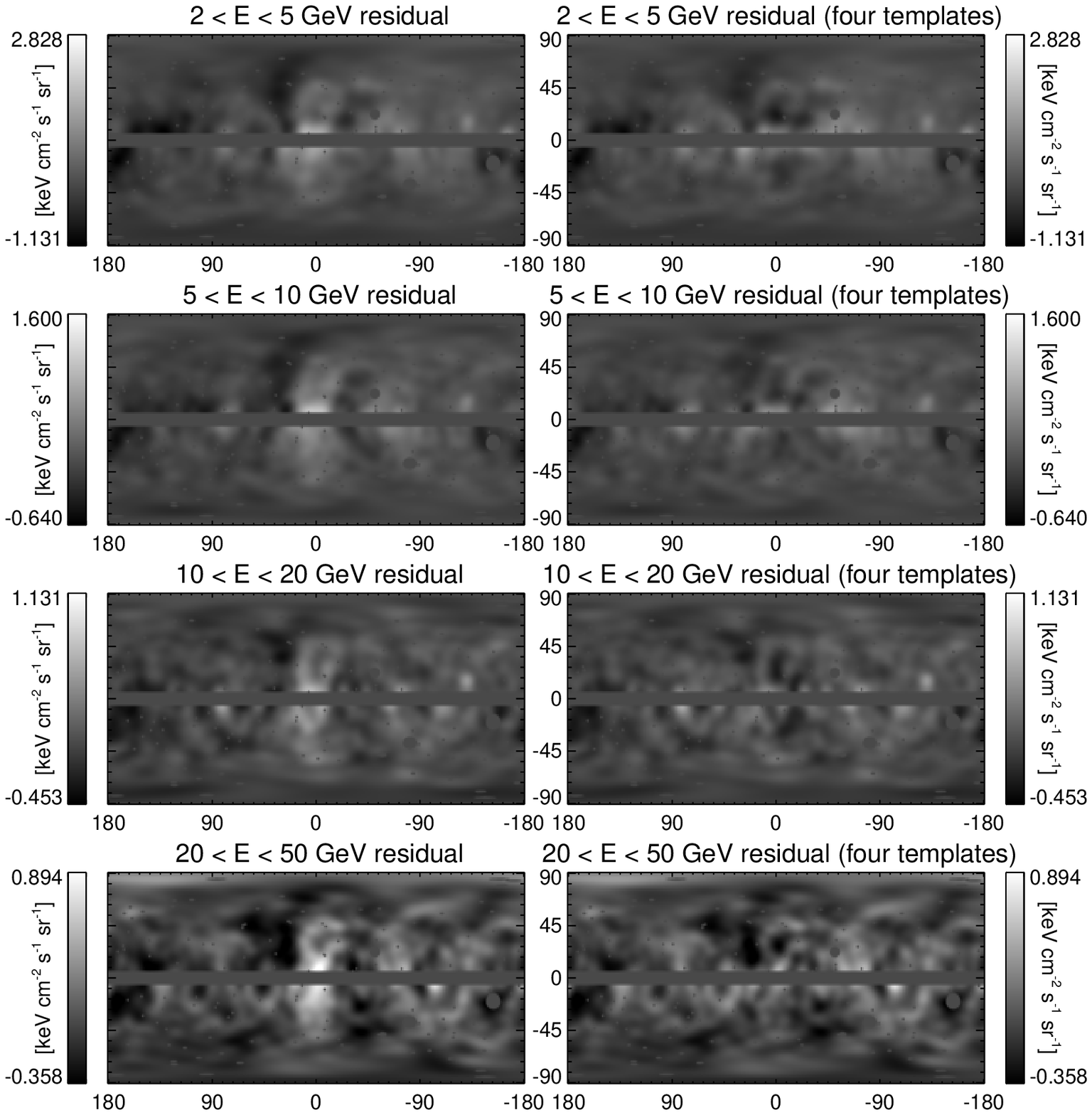}
}
\caption{ 
  The same as \reffig{fermi_template_residual} but for a FWHM
  smoothing of $10\degree$.  Smoothing the residuals at this scale
  demonstrates that the haze is a robust feature and, in particular,
  is not the consequence of single photon fluctuations at high
  energies.
}
\label{fig:fermi_template_residual_smth}
\epm

\bpm
\centerline{
  \includegraphics[width=0.9\textwidth]{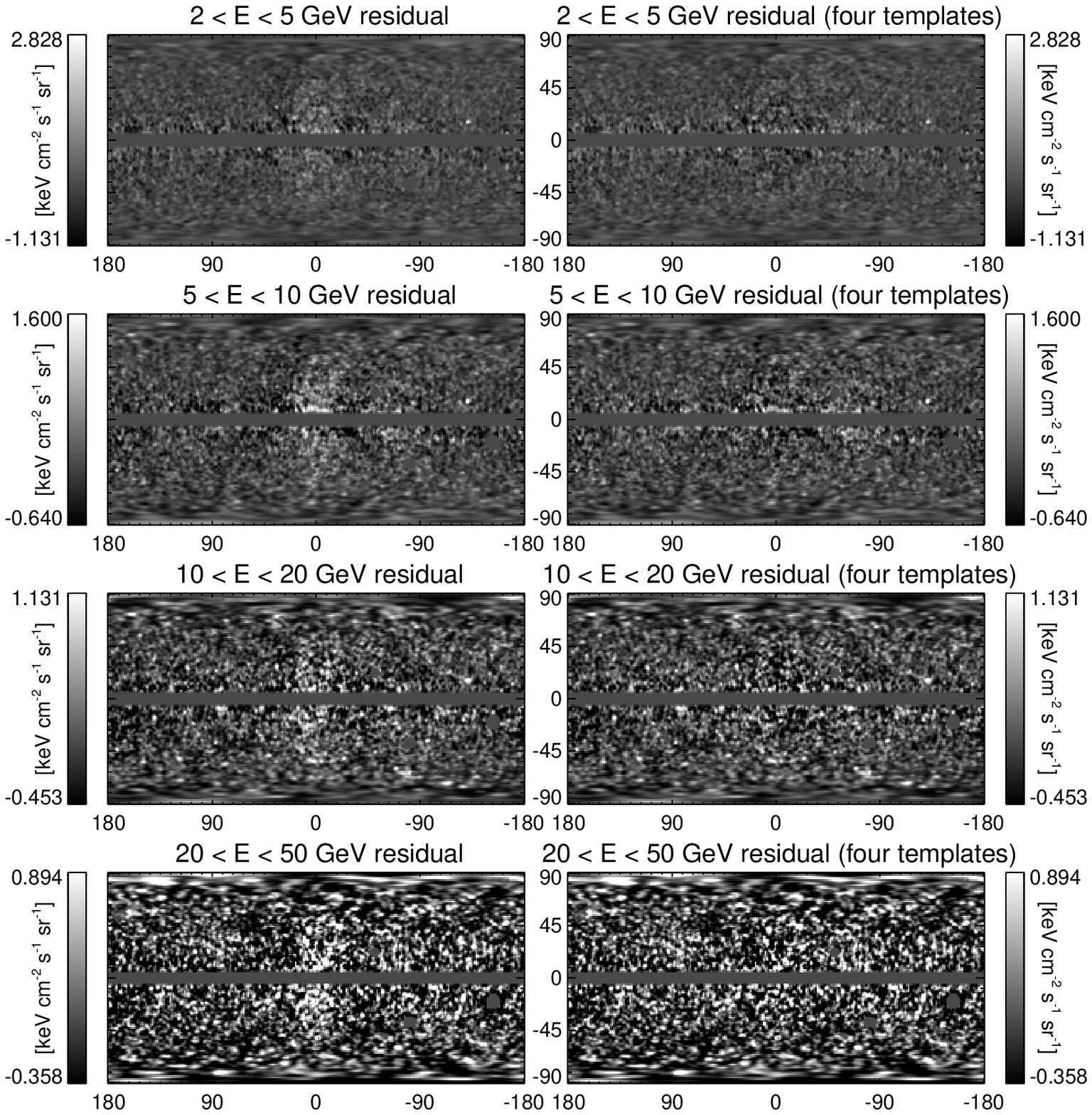}
}
\caption{ 
  The same as \reffig{fermi_template_residual} but using the 1-2 GeV
  map instead of the SFD dust map to trace $\pi^0$ emission.  The
  clear haze residual seen, particularly at high energies, indicates
  that the gamma-ray haze is not due to shortcomings of the SFD
  template resulting from variations in the proton cosmic ray density.
  These variations, as well as line of sight gas density effects, are
  automatically included in the 1-2 GeV template.
}
\label{fig:fermi_template_residual_loE}
\epm

In \reffig{intensity_vs_b} we show the haze amplitude (residual maps
from the Type 3 fit plus the correlation-coefficient-weighted
bivariate Gaussian template) as a function of Galactic latitude and
for the longitudinal bin: $-15\degree < \ell < 15\degree$.  Although
the data are noisy, the figure shows that the \emph{Fermi} haze dies
off by roughly $b \sim 50\degree$ in all energy bands.  For
comparison, we show the same plot for just the residual map.  The
figure also shows the amplitude as a function of longitude for the
latitudinal bin $-20\degree < b < -10\degree$.  The more rapid fall
off of the haze emission with $\ell$ compared to $b$ indicates a haze
morphology elongated in the $b$ direction.

\bpm
\centerline{
  \includegraphics[width=0.9\textwidth]{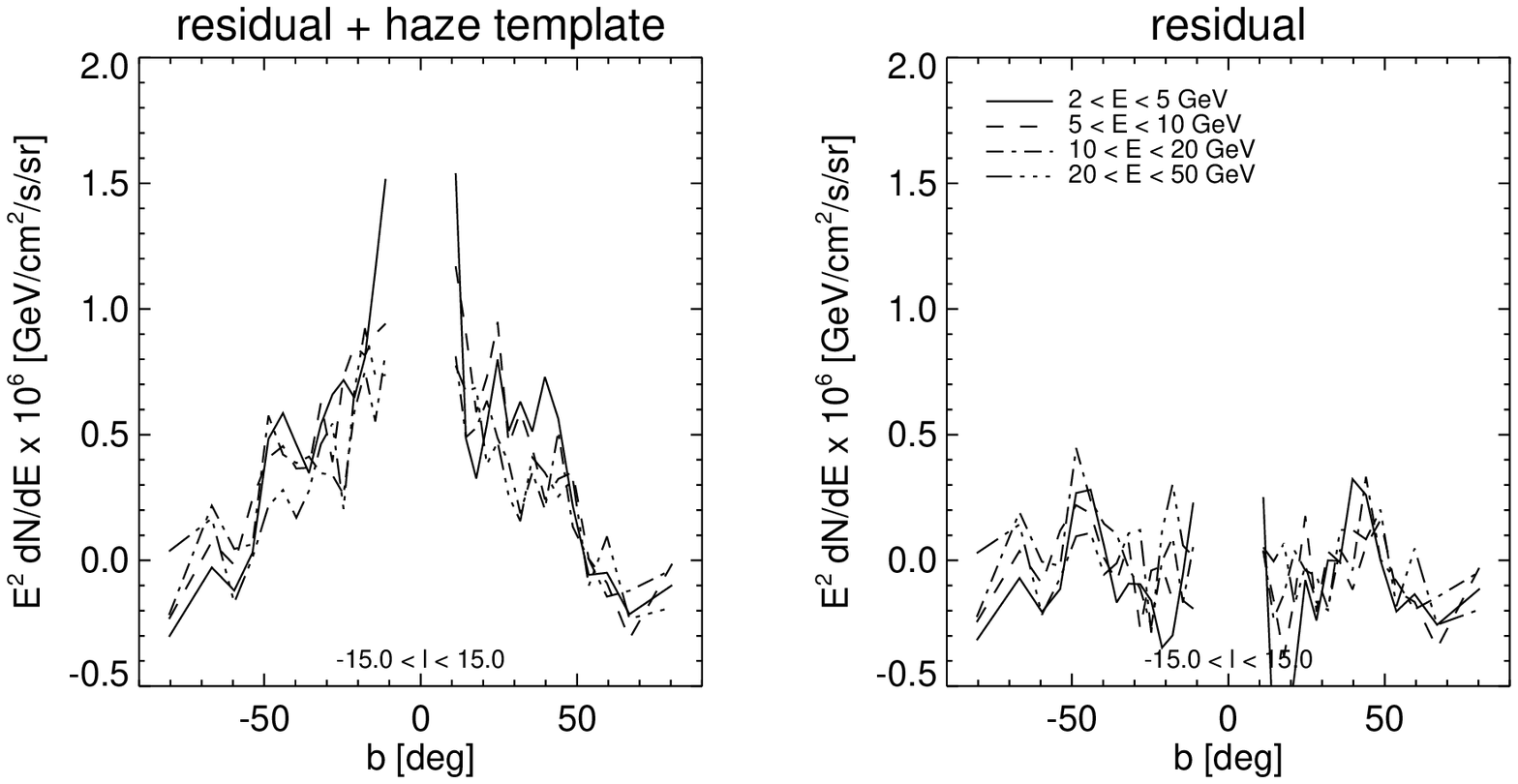}
}
\centerline{
  \includegraphics[width=0.9\textwidth]{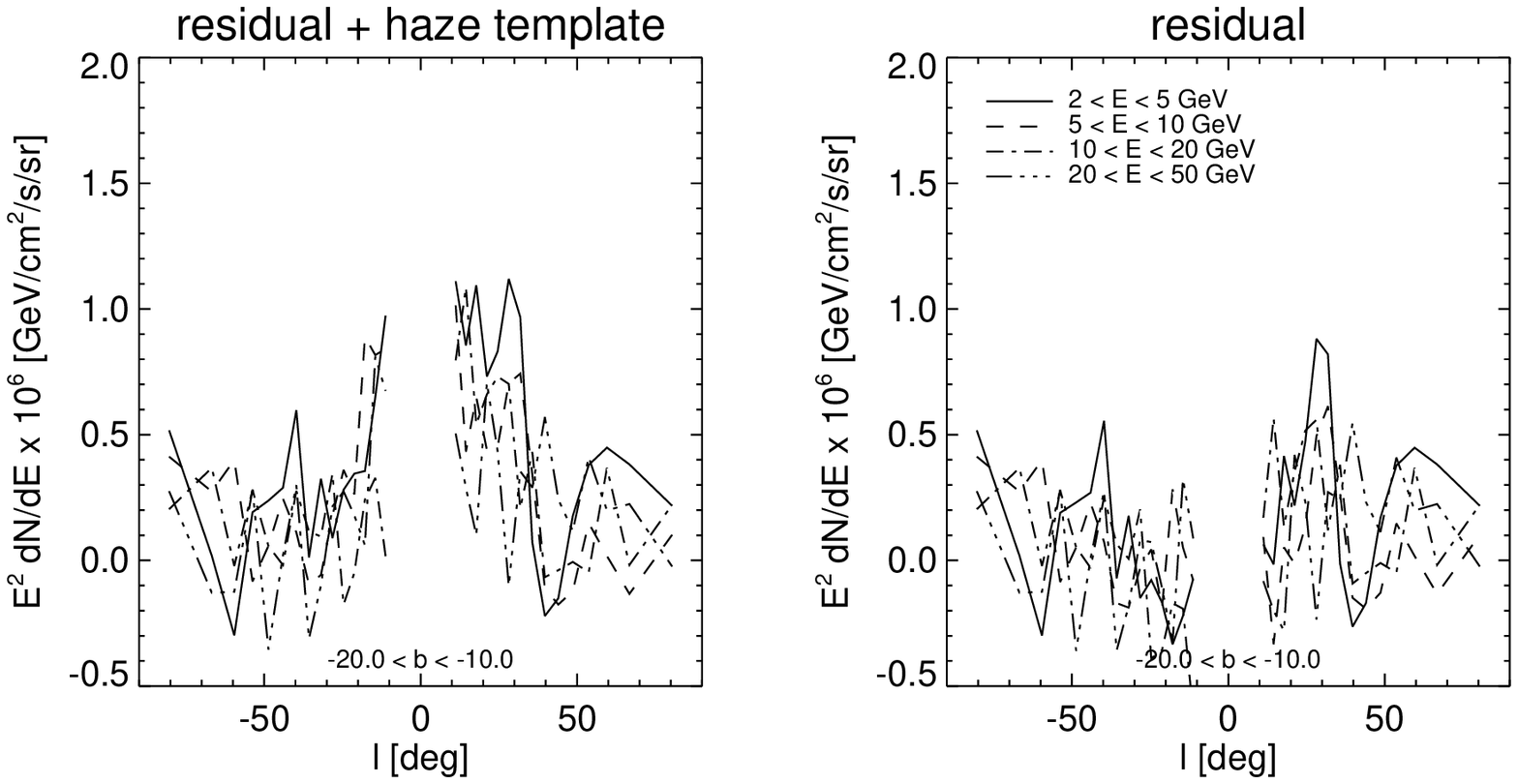}
}
\caption{ 
  \emph{Top left:} The haze amplitude (Type 3 residual map plus haze
  template) as a function of Galactic latitude for four different
  energy ranges.  The data are binned in steps of 0.055 in $\sin b$,
  i.e. roughly $4\degree$ bins near the plane.  The plot shows that
  haze amplitude has roughly the same fall off with $b$
  ($\sim40\degree$) in all energy ranges.  \emph{Top right:} The same
  but for the Type 3 residual map.  \emph{Bottom left and right:} the
  same as the top left and right, but as a function of $\ell$ for a
  fixed range of $b$.  The haze profile clearly falls off more quickly
  with $\ell$ than with $b$ indicating a profile elongated in the $b$
  direction.
}
\label{fig:intensity_vs_b}
\epm

There are two important features to note about Figures
\ref{fig:fermi_sub_sfd} through \ref{fig:fermi_template_residual}.
First, the IC haze has a spectrum which is \emph{harder} than the
other IC in the Galaxy.  This is evidenced by the fact that using the
1-2 GeV \emph{Fermi} map as a template removes the IC emission from
disk electrons but the haze IC fades more slowly with energy.
Second, the IC haze is morphologically distinct from either the
$\pi^0$ emission or IC and bremsstrahlung from the disk electrons.
These two facts taken together strongly suggested that the electrons
responsible for the microwave and gamma-ray haze are from a
\emph{separate} component with a \emph{harder spectrum} than SN
shock-accelerated electrons.

\subsection{\emph{Fermi} Galactic Diffuse Emission Model}
\label{sec:fermidiffuse}
So far, we have only done template fits of maps of data to other maps of
data.  Using the SFD dust map as a tracer of $\pi^0$ emission is
equivalent to assuming that the proton CR spectrum and density are
spatially uniform, and the dust/gas ratio is constant.  Using the Haslam
map to estimate IC emission is equivalent to assuming that the B-field
and ISRF have similar spatial variation, and neglecting the anisotropies
in both IC and synchrotron emission.  Consequently, there have been
concerns \citep{Linden:2010ea} that this template-based analysis might
introduce spurious large-scale residuals.
To address these concerns, we investigate whether the \emph{Fermi}
diffuse model contains a structure similar to the haze emission.

The \emph{Fermi} team provides a model of diffuse gamma-ray emission
consisting of maps sampled at 30 energy bins from 50 MeV to
100 GeV.\footnote{ The
  background models can be downloaded
  from\\ \texttt{http://fermi.gsfc.nasa.gov/ssc/data/access/lat}.}
These maps are based on template fits to the gamma-ray data and also
include an IC component generated by the GALPROP cosmic ray
propagation code.  

The difference between the model and data is shown in \reffig{diffmod}.
We have interpolated the model to the energy ranges of interest and
performed the simple one-template fit to the data (analogous to our Type 1 and 2
fits described above).  This fit allows for any error in the
normalization of the diffuse model; the fit coefficients are within a
few percent of unity in every case.

At each energy range, the haze is clearly visible in the residual maps.
GALPROP uses the standard inhomogeneous ISRF model
\citep{Porter:2005qx}, making it unlikely that the observed residual is 
due to the expected spatial variation of the ISRF.
Furthermore, the diffuse model includes a
disk-like injection of primary electrons and estimates for the line of
sight density variations of the interstellar medium.  The fact that
the haze residual remains suggests that, given the propagation model
included in GALPROP, disk-like sources will not produce the observed
IC emission, nor will expected variations
in the gas density and proton CR density along the line of sight.
\bpm
\centerline{
  \includegraphics[width=0.9\textwidth]{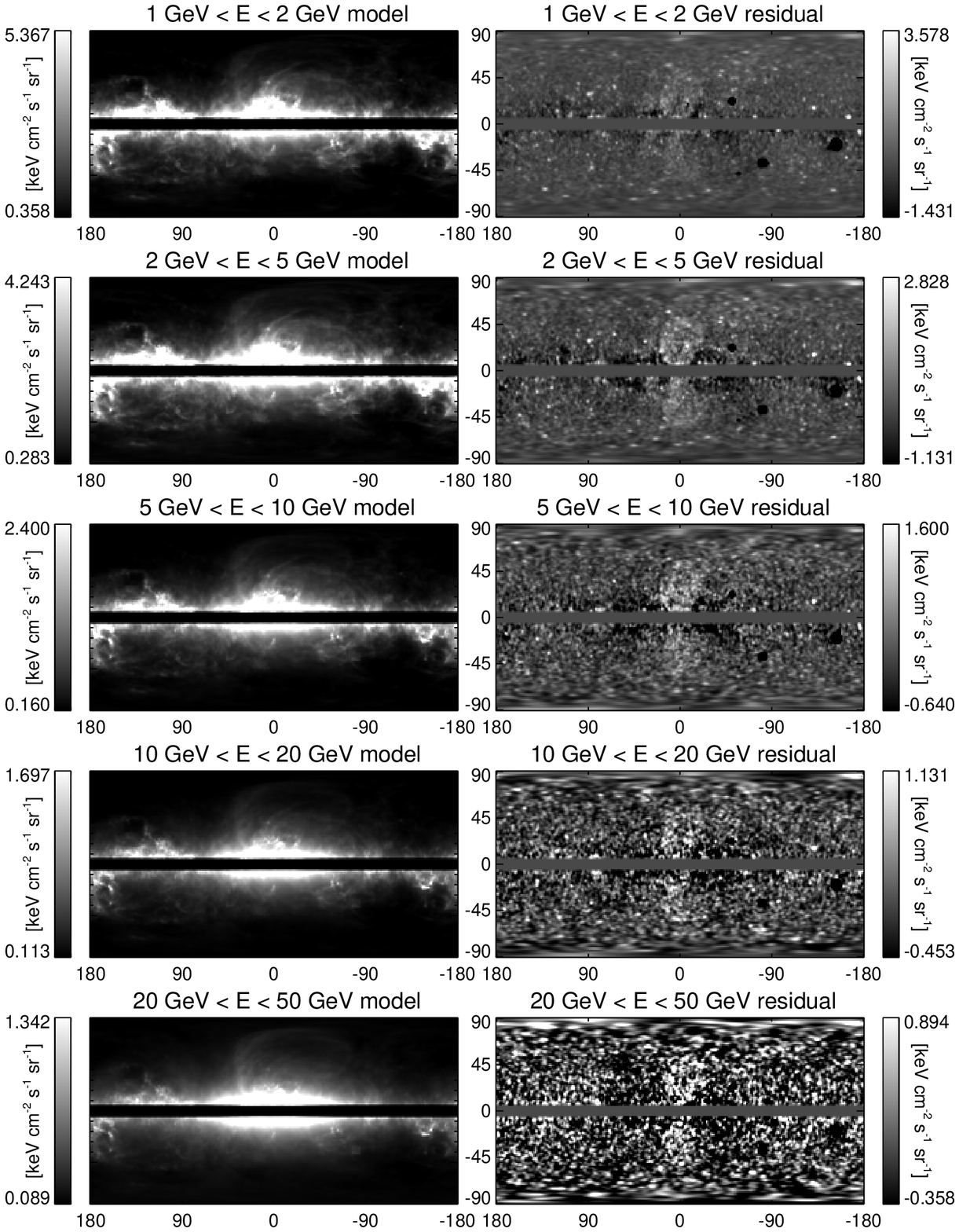}
}
\caption{ 
  The same as \reffig{fermi_sub_sfd} but using the official
  \emph{Fermi} team Galactic diffuse model.  The strong haze residual
  in the right hand panels shows that the haze is \emph{not} included
  in the model.  Since the IC emission in the model was obtained with
  GALPROP, this indicates that variation in the ISRF with position in
  the Galaxy cannot account for the haze emission.  In addition, the
  haze morphology is not reproduced by disk-like injection of
  electrons, nor by the cosmic ray propagation model employed by
  GALPROP.
}
\label{fig:diffmod}
\epm

\subsection{Spectra}
\label{sec:spectra}

While the morphology of the gamma-ray haze is indicative of IC
emission from the microwave haze electrons, we now attempt to estimate
the spectrum of this emission.  This is difficult for several reasons.
First, $\pi^0$ emission is dominant (or nearly so) at most energies in
\emph{Fermi}'s energy range.  Thus, in a given region, we must
estimate the spectral shape and amplitude of the $\pi^0$ emission in
order to subtract it from the total.  Second, the total number of
photons measured by \emph{Fermi} decreases rapidly with increasing
energy.  For example, in the inner Galaxy ($|\ell| \leq 30$, $|b| \leq
5$), there are only $\sim$4300 photons between 10 and 100 GeV,
resulting in substantial Poisson errors.  Lastly, there is also
significant isotropic background due to extra-galactic emission and
charged particle contamination, including heavy nuclei at high
energies.  We will show that the isotropic backgrounds are manageable,
and present two types of spectra: template coefficients $c_{\rm T}(E)$
and total fluxes $dN/dE$.

\subsection{Isotropic Background}
\label{sec:background}

The \emph{Fermi} data contain gamma-rays from an unresolved
extragalactic signal with $dN/dE \sim E^{-2.45}$ \citep{ACKERMANNTALK}
as well as particle contamination.  We make no attempt to separate the
extragalactic gamma-ray signal from the particle contamination.
Instead we measure the spectrum of this (nearly) isotropic background
in 8 regions at high latitude and test the assumption of
isotropy. Specifically, we take combinations of longitude ranges of
$-180^\circ < \ell < -90^\circ$, $-90^\circ < \ell < 0^\circ$,
$0^\circ < \ell < 90^\circ$ and $90^\circ < \ell < 180^\circ$ together
with latitude ranges of $-90^\circ < b< -60^\circ$ and $60^\circ < b<
90^\circ$ for the $4\times2=8$ regions.  The regions are chosen to be
far from the plane to avoid contamination from Galactic emissions. We
use the point-source masked maps described above, binned in 12
energy bins from 0.3 - 300 GeV.  The energy bins, mean background,
etc.\ are given in Table \ref{tbl:background} and the 8 spectra, along
with the inverse-variance weighted mean, are plotted in
\reffig{background}.  We note that the error bars are the standard
deviation of the 8 regions, \emph{not} the standard deviation of the
mean.  To be conservative, we use this standard deviation as the
uncertainty in the background in the remaining stages of the analysis.
Future work by the \emph{Fermi} team to understand both the particle contamination and the
gamma-ray sky will likely reduce this error in the background
substantially.

\begin{deluxetable}{|c|c|c|c|}
  \tablehead{ E range & Energy & background & $\chi_\nu^2$ \\
               (GeV)  & (GeV) & ($\times 10^{-6}$GeV/cm$^2$/s/sr) &  }
  \startdata
$   0.3-  0.5$ & $   0.4 $ & $ 2.235 \pm 0.187 $ & $  86.467 $ \\
$   0.5-  0.9$ & $   0.7 $ & $ 1.901 \pm 0.190 $ & $  76.812 $ \\
$   0.9-  1.7$ & $   1.3 $ & $ 1.657 \pm 0.181 $ & $  51.885 $ \\
$   1.7-  3.0$ & $   2.2 $ & $ 1.340 \pm 0.157 $ & $  27.169 $ \\
$   3.0-  5.3$ & $   4.0 $ & $ 1.093 \pm 0.119 $ & $  10.981 $ \\
$   5.3-  9.5$ & $   7.1 $ & $ 0.886 \pm 0.069 $ & $   2.720 $ \\
$   9.5- 16.9$ & $  12.7 $ & $ 0.687 \pm 0.098 $ & $   4.190 $ \\
$  16.9- 30.0$ & $  22.5 $ & $ 0.632 \pm 0.068 $ & $   1.357 $ \\
$  30.0- 53.3$ & $  40.0 $ & $ 0.695 \pm 0.078 $ & $   0.936 $ \\
$  53.3- 94.9$ & $  71.1 $ & $ 0.624 \pm 0.125 $ & $   1.183 $ \\
$  94.9-168.7$ & $ 126.5 $ & $ 0.704 \pm 0.199 $ & $   1.594 $ \\
$ 168.7-300.0$ & $ 225.0 $ & $ 0.420 \pm 0.205 $ & $   1.949 $
  \enddata
  \tablecomments{The background is tabulated in each energy bin for 8
    polar ($|b|>60\degree$) regions after point source masking.  The
    uncertainty is the RMS of the 8 regions, \emph{not} the
    uncertainty in the mean, which is $\sqrt{7}$ times smaller.  For
    each region we also compute the Poisson uncertainty, $\sigma_p$,
    which for low $E$ is much smaller.  To test the significance of 
    background variation, we compute $\chi_\nu^2 = \langle \sigma^2 /
    \sigma_p^2 \rangle$, 
    averaging over the 8 regions.  There is no
    strong indication of anisotropy at high latitude for $E>5$ GeV.
}\label{tbl:background}
\end{deluxetable}

\bp
  \includegraphics[width=0.49\textwidth]{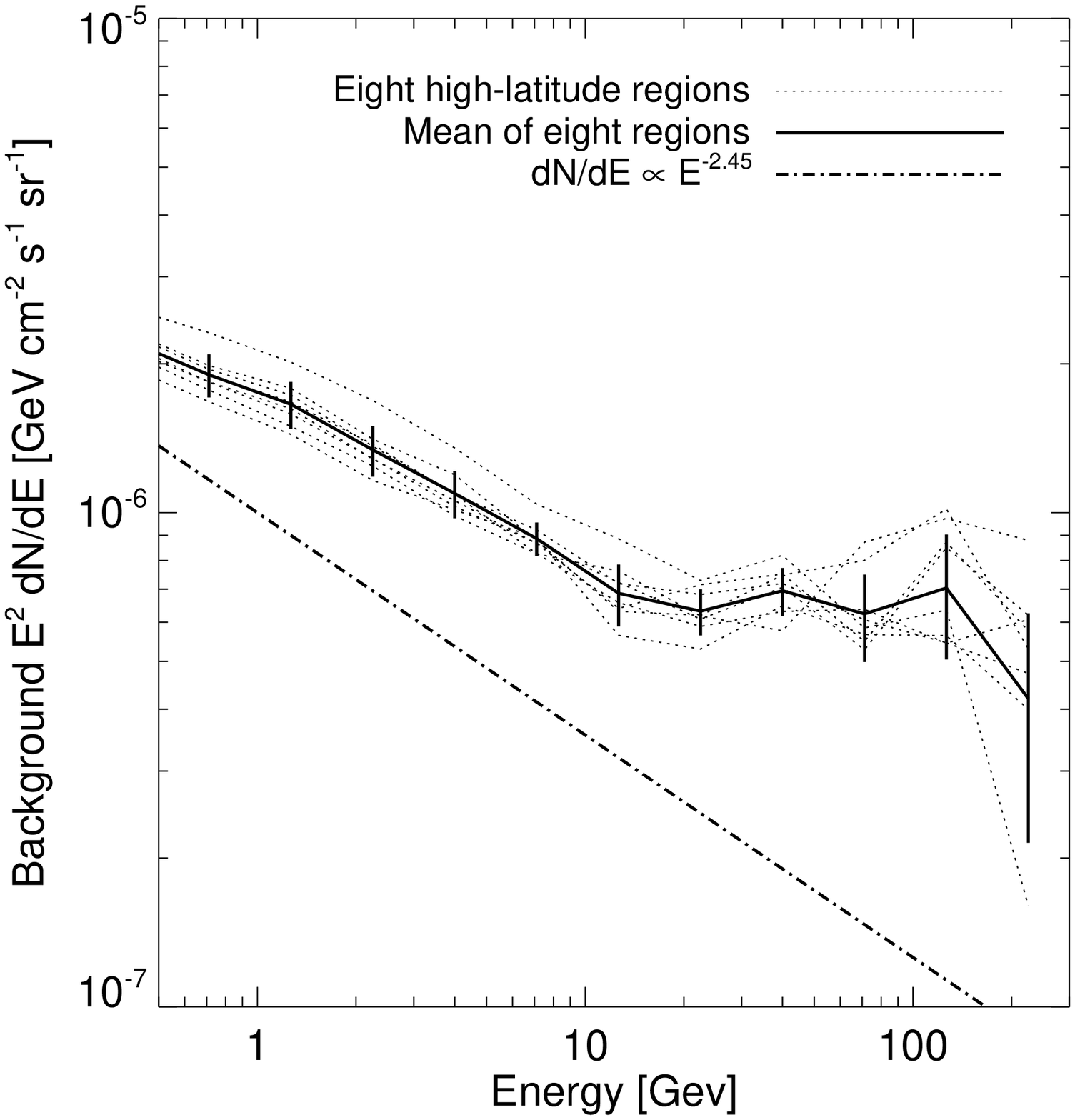}
\caption{
  Background spectrum, including gamma-rays and particle backgrounds.
  The background is tabulated in each energy bin for 8 high-latitude
  polar ($|b|>60\degree$) regions after point source removal (dotted
  gray lines).  The mean of the 8 regions is shown with a solid black
  line and the uncertainty is the RMS of the 8 regions, \emph{not} the
  uncertainty in the mean, which is $\sqrt{7}$ times smaller.  For
  reference, a $dN/dE \propto E^{-2.45}$ model for the unresolved
  background is shown with a dot-dashed line.
}
\label{fig:background}
\ep

\bpm
  \includegraphics[width=0.49\textwidth]{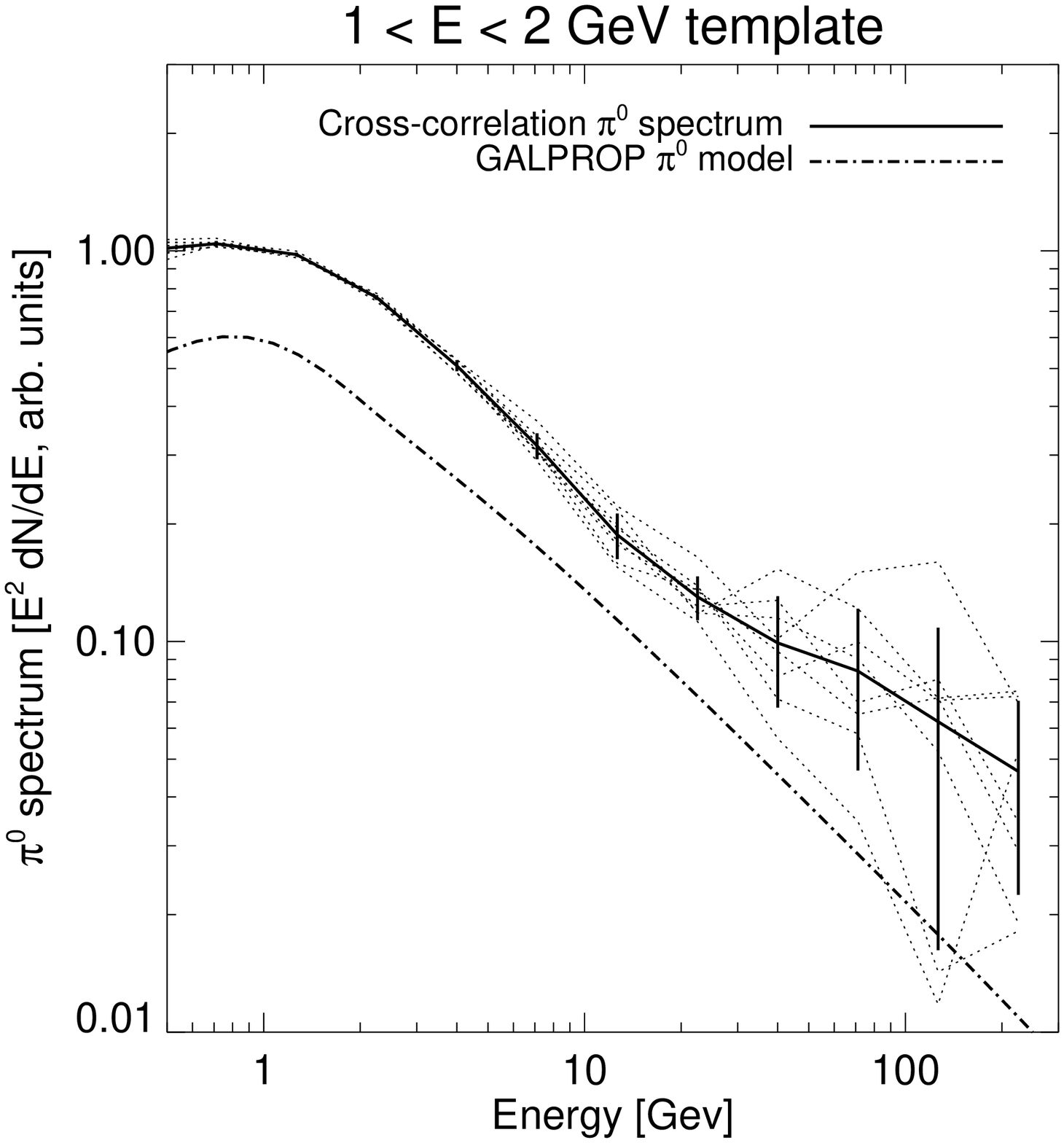}
  \includegraphics[width=0.49\textwidth]{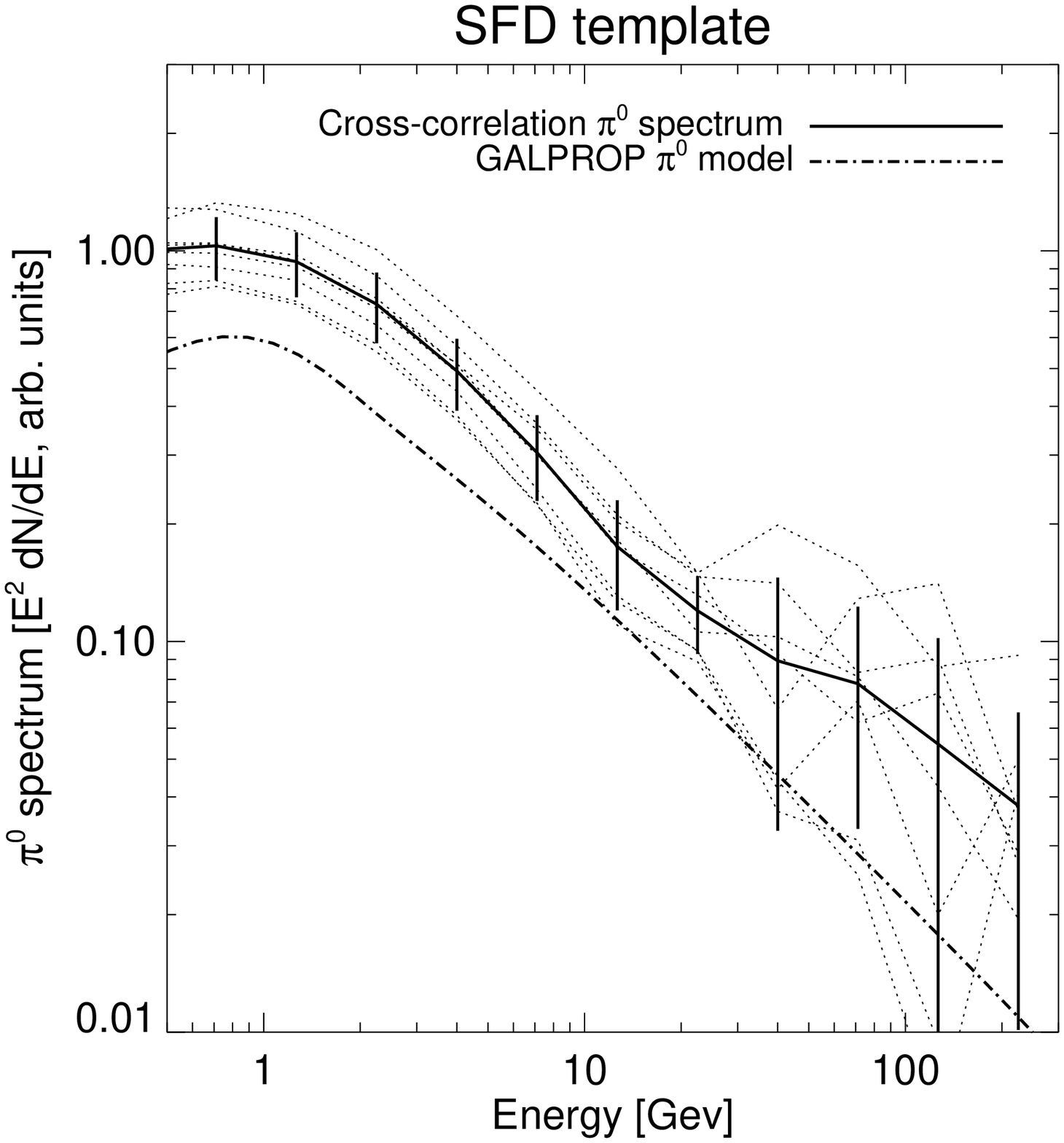}
\caption{ 
  \emph{Left:} The $\pi^0$ spectrum derived using the
  cross-correlation technique defined in \refsec{residualmaps} and
  with the \emph{Fermi} 1-2 GeV map as a $\pi^0$ template.  The dotted
  Gray lines are the cross-correlation spectra in several different
  regions of the sky and the solid black line is the mean of those
  spectra.  Error bars on the cross-correlation spectrum are defined
  as the variance in the values for the different regions.  The
  dot-dashed line is the $\pi^0$ spectrum output from a GALPROP model
  (shown here with arbitrary normalization).  \emph{Right:} The same
  but using the SFD dust map as a template.
}
\label{fig:pi0_spectrum}
\epm

\subsection{Template-Correlated Spectra}
\label{sec:temp_spec}

\reffig{pi0_spectrum} shows $c_{\rm T}(E) \times \mean{T}$ for the two
templates and regions 1-7 used in the Type 1 and Type 2 fits along
with the model $\pi^0$ spectrum from GALPROP, which uses the 
\cite{Blattnig:2000zf} parameterizations for pion production.  It is clear from the
figure that the cross-correlation technique produces $\pi^0$ spectra
that are remarkably similar to the model spectrum at low energies,
while at high energies the cross-correlation spectrum is slightly
higher than the model spectrum.  This could be due to a number of
reasons such as non-zero spatial correlation between the templates and
the harder spectrum haze IC, contamination from background events
like heavy nuclei, or uncertainties in the $\pi^0$ emission model.  Of
these, the first is most likely since the cross-correlation between
the templates and a nearly isotropic background is likely small and
since the spectrum of $\pi^0$ gammas is quite well
known.

Template-correlated spectra for the Type 3 template fit are shown in
\reffig{template_spec}.  Here the correlation coefficients are
weighted by the mean of each template in the ``haze'' region (see
\reftbl{roi}).  As shown in the figure, the spectra for the SFD and
Haslam maps reasonably match the model expectations in that
region\footnote{The GALPROP model here was tuned to match locally
  measured protons and anti-protons as well as locally measured
  electrons at $\sim20$-30 GeV.}.  That is, the SFD-correlated emission
roughly follows the model $\pi^0$ spectrum while the Haslam-correlated
spectrum resembles a combination of IC and bremsstrahlung emission.
However, the \emph{haze}-correlated emission is clearly significantly
harder than either of these components.  This fact coupled with the
distinct spatial morphology of the haze indicates that the IC haze is
generated by a \emph{separate} electron component.

\bpm
\centerline{
  \includegraphics[width=0.98\textwidth]{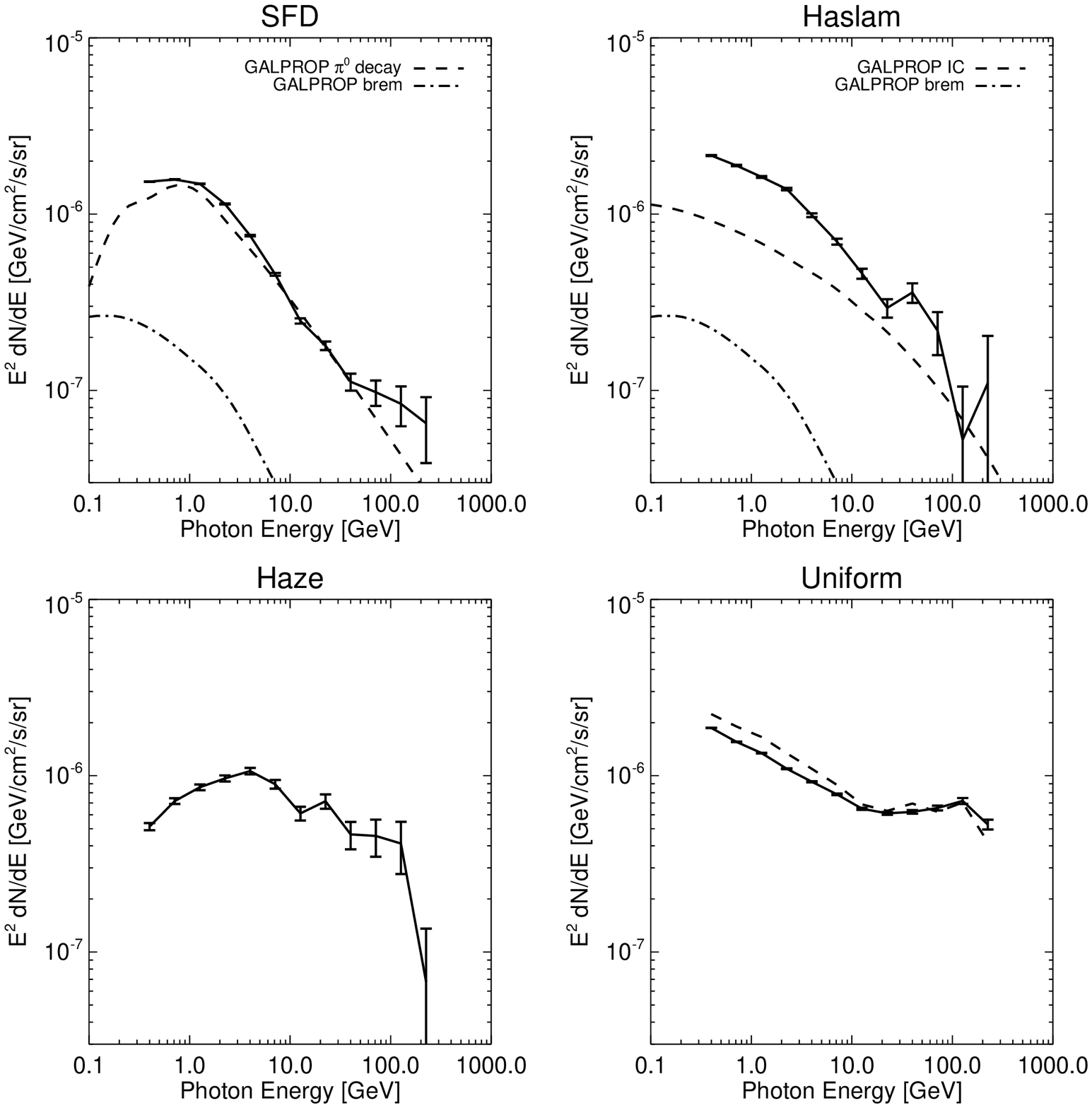}
}
\caption{
  Correlation coefficients for the templates used in the Type 3
  template fit (see \refsec{residualmaps}).  \emph{Upper left:}
  SFD-correlated spectrum which roughly traces $\pi^0$ emission.
  \emph{Upper right:} Haslam-correlated emission which traces the soft IC and bremsstrahlung component.  
  \emph{Lower left:} haze
  template-correlated emission.  This component has a notably harder
  spectrum than both the SFD- and Haslam-correlated spectra or their
  model shapes (dashed lines, cf., \reffig{galprop_example}),
  indicating a separate component.  The dashed lines in the upper two
  panels are a GALPROP estimate of $\pi^0$ decay (left) and
  IC emission (right). The GALPROP estimate of bremsstrahlung emission is shown in both panels. \emph{Lower right:} the
  uniform template-correlated spectrum which traces the isotropic
  background.  Here the dashed line is the result from
  \reffig{background}.  This high latitude estimate is higher than the
  uniform template estimate likely because the $\pi^0$ emission is
  non-zero at high latitudes and leaks into our measured background.
  This is less of an issue for the uniform template which uses the
  morphological (i.e., uniform) information.
}
\label{fig:template_spec}
\epm

\subsection{Total Intensity Spectra}
\label{sec:inten_spec}

While the template-correlated spectra and residual maps are useful for
identifying separate components, for the purposes of comparing the map
intensities to a model for the physical mechanisms, we now generate
total intensity spectra in several regions of interest.  We define
three key regions: a ``haze region'' south of the GC\footnote{The
  region south of the GC is greatest interest for studying the
  microwave and gamma-ray haze.  This is because, north of the GC, the
  microwave maps from WMAP include bright free-free emission from
  $\zeta$ Oph, and spinning dust emission from $\rho$ Oph, both of
  which are bright enough to leave substantial residuals, even though
  they are relatively well subtracted by the template fitting described in
  \cite{Dobler:2008ww}.  South of the Galactic center there are some
  small dust and gas features that should provide some signal in the
  gamma-ray map, but the situation is much simpler.}, the ``four
corners'' region from \cite{Cholis:2009gv}, and a Galactic plane
region used by \cite{TROYTALK}.

\begin{deluxetable}{|c|c|c|}
  \tablehead{ Region & $\ell$ range & b range}
  \startdata
    Inner Galaxy & $|\ell|<30\degree$        & $|b| < 5\degree$ \\
    Haze         & $|\ell|<15\degree$        & $-30 < b < -10\degree$ \\
    4 Corners    & $5 < |\ell| < 10\degree$  & $5 < |b| < 10\degree$
  \enddata
\tablecomments{
  Regions for which the total gamma-ray intensity is evaluated (see
  also \reffig{spectra}).
}\label{tbl:roi}
\end{deluxetable}

The spectrum
of each region is shown (\reffig{spectra}), with the background spectrum from
\reffig{background} subtracted from the other spectra to remove any
isotropic component.  The inner Galaxy region clearly shows the low
energy behavior characteristic of $\pi^0$ emission \citep[though it is
  important to note that there may be significant emission from
  \emph{unresolved} point sources such as pulsars in this region which
  have a similar spectral shape; see \refapp{pointsources} and
][]{Abdo:2009ax}.  At high energies however, there is a significant
excess which is not expected for $\pi^0$ emission
(cf., \reffig{galprop_example}).  The spectrum of this excess is similar
to the spectrum derived for the haze template from
\reffig{template_spec} and is consistent with an IC signal from a
hard electron population with energies $>10$ GeV.  The evidence for
this excess is more pronounced in the haze and four corners region.
Lastly, we take the template estimate of $\pi^0$ emission in the haze
region from the Type 3 fits and subtract it from the total emission.
This reveals two clear features: a bump centered on roughly 1-2 GeV
that is likely due to either bremsstrahlung from a low energy electron
component or emission from unresolved pulsars, and a hard tail
above $\sim 10$ GeV that is the IC signal from the haze
electrons.

\bp
\centerline{
  \includegraphics[width=0.49\textwidth]{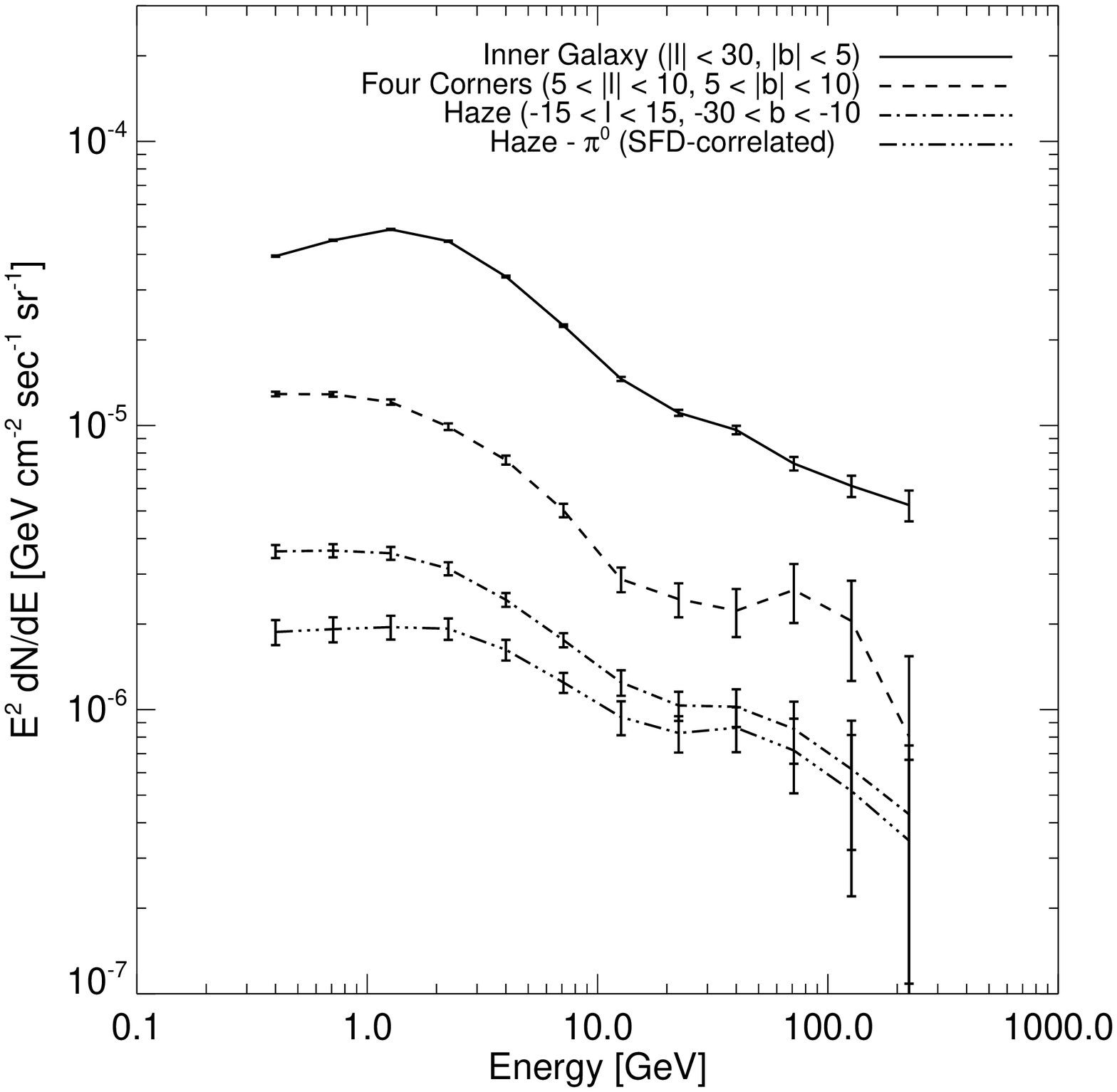}
}
\caption{ 
  Total intensity spectra in three regions of interest (solid, dashed,
  dot-dashed; see \reftbl{roi}).  All three show the characteristic
  $\pi^0$ spectrum at low energies as well as an excess at
  higher energies with roughly the same spectrum as the haze spectrum
  derived from the Type 3 template fits.  Furthermore, subtracting the
  template $\pi^0$ spectrum in the haze region (dot-dot-dot-dashed)
  yields two features, one centered on $\sim2$ GeV and one
  centered on $\sim50$ GeV.  The lower energy feature is likely either
  bremsstrahlung from a low energy electron population (like those
  accelerated by SN shocks) or emission from unresolved pulsars.  The
  higher energy tail is the IC signal from the haze electrons.
}
\label{fig:spectra}
\ep

\subsection{Comparison to the Microwave Haze}
\label{sec:comparehaze}

Both the morphology and the relatively hard spectrum of the gamma-ray haze motivate a common physical origin with the WMAP haze. We now provide a simple estimate of the microwave and gamma-ray signals from a population of hard electrons in the inner Galaxy, to demonstrate that the magnitudes and spectral indices of the two signals are consistent for reasonable parameter values.

We consider a steady-state electron spectrum described by a power law, $dN/dE \propto E^{-\alpha}$, with a high-energy cutoff at 1 TeV (here the cutoff is implemented as a step function, not an exponential fall-off; of course this is only an approximation to the true spectrum). This choice is motivated by the local measurement of the cosmic ray electron spectrum by \Fermi \citep{Abdo:2009zk}. We consider a region $\sim 2$ kpc above the Galactic center, as an example (and since both hazes are reasonably well measured there), and employ the model for the ISRF used in GALPROP \citep{Porter:2005qx} at 2 kpc above the center. We normalize the synchrotron to the approximate value measured by WMAP in the 23 GHz K-band \citep{Hooper:2007kb}, $\sim 15^\circ$ below the Galactic plane, and compute the corresponding synchrotron and IC spectra. The WMAP Haze was estimated to have a spectrum $I_\nu \propto \nu^{-\beta}$, $\beta = 0.39-0.67$ \citep{Dobler:2008ww}, corresponding approximately to an electron spectral index of $\alpha \approx 1.8-2.4$;  Figure \ref{fig:comparehaze} shows our results for a magnetic field of 10 $\mu$G, and electron spectral indices $\alpha = 2-3$. This field strength is appropriate for an exponential model for the Galactic $B$-field intensity, $|B| = |B_0| e^{-z/z_s}$, with $B_0 \approx 20-30 \mu$G (which seems reasonable; see e.g. \cite{Ferriere:2009dh} and references therein) and scale height $z_s \approx 2$ kpc (the default value in GALPROP). We find good agreement in the case of $\alpha \approx 2-2.5$, consistent with the spectrum of the WMAP Haze.

\bp
\centerline{
  \includegraphics[width=0.49\textwidth]{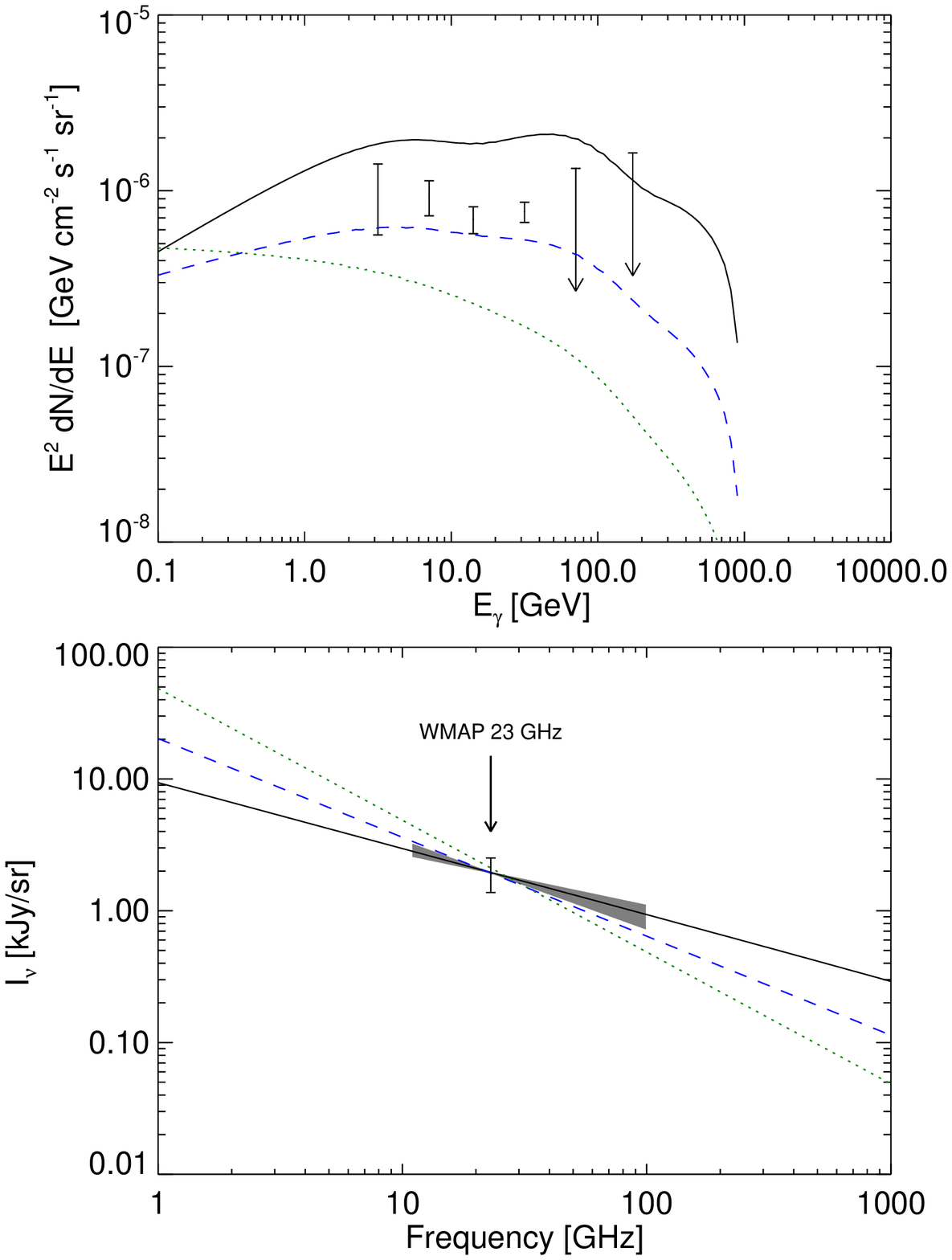}
}
\caption{The estimated spectrum of IC gamma rays (\emph{upper panel}) and synchrotron radiation (\emph{lower panel}) originating from a hard electron spectrum along a line of sight 2 kpc above the Galactic center (i.e. $b \approx 15\degree$). The steady-state electron spectrum is taken to be a power law, $dN/dE \propto E^{-\alpha}$, with index $\alpha = 2$ (\emph{solid}), 2.5 (\emph{dashed}) and 3 (\emph{dotted}). In all cases the spectrum has a cutoff at 1 TeV. The interstellar radiation field model is taken from GALPROP, and the magnetic field is set to be 10 $\mu$G. The data points in the upper panel show the magnitude of the \Fermi~ Haze averaged over $|b|=10-18$, for $|l| < 15$, as a function of energy, taken from Figure \ref{fig:intensity_vs_b}. The highest two bins contain 3 $\sigma$ upper limits rather than data points with $1 \sigma$ error bars, due to the large uncertainties in the Haze amplitude at those energies. The data point in the lower panel shows the magnitude of the WMAP Haze averaged over $b=-10$ to $-18$, for $|l| < 10$, in the 23 GHz K-band (the overall normalization is chosen to fit this value), and the gray area indicates the range of synchrotron spectral indices allowed for the WMAP Haze by \cite{Dobler:2008ww}. 
}
\label{fig:comparehaze}
\ep

\subsection{Comments on Haze Morphology}
\label{sec:haze morphology}

Although a detailed analysis of the possible sources of the
\emph{Fermi} haze are beyond the scope of this paper, a few simple
comments are in order. First, the profile is not well described by a
disk source. While quantifying this is a subtle task, the success of
the template makes this point clear. There are many possibilities to
explain the oblong shape, should that persist in future data. For
instance, AGN jets and triaxial DM profiles could both produce signals
of this shape. Even approximately spherical injection could yield such
a signal, should the diffusion be considerably anisotropic. That said,
while one might have attempted to invoke, e.g., rapid and significant
longitudinal variation of the magnetic field to explain the microwave
haze with a disk-like electron injection profile, such approaches are
no longer tenable (and moreover already had significant tension with
an understanding of the Haslam synchrotron maps). The presence of this
feature in both gamma rays and microwaves demonstrates that {\em the
  electrons themselves} do not follow a disk-like injection profile.
Furthermore, we emphasize that the angular scale of the haze is very
large, roughly 25-40$\degree$ in $b$.  This region of the sky is far
from the Galactic disk or the very GC where complications from the
central black hole, other point sources, or disk-like pulsars could
complicate the analysis.

\section{Discussion/Conclusions}

We have presented full sky maps generated from photon events in the
first year data release of the \emph{Fermi Gamma-Ray Space Telescope}
(see Appendix \ref{sec:datarelease} for data processing details).

Using a template fitting technique, we have approximated both the
spectrum and morphology of the well known gamma-ray emission
components at \emph{Fermi} energies.  The SFD dust map was used to
trace the $\pi^0$ decay gammas generated by collisions of cosmic ray
protons with the ISM, while the Haslam 408 MHz map was used to trace
inverse Compton (IC) scattered photons from interactions of supernova
shock-accelerated electrons ($\sim 1-10$ GeV) with the interstellar
radiation field (ISRF).  Bremsstrahlung radiation, generated by
interactions of these electrons with the ISM, should be approximately
traced by some combination of these two maps.  Although our template
fitting technique is subject to significant uncertainties due to
uncertain line of sight gas and CR distributions, a robust positive
residual has been identified.

This excess diffuse emission is centered on the Galactic center, and
can be parameterized by a simple two-dimensional Gaussian template
$(\sigma_\ell=15\degree, \sigma_b=25\degree)$. The template-correlated
spectrum of this emission is significantly \emph{harder} than either
$\pi^0$ emission or IC from softer electrons, whose fitted spectra
agree well with models.  This harder spectrum coupled with the
distinct spatial morphology of the gamma-ray and microwave haze are
evidence that these electrons originate from a \emph{separate
  component} than the softer SN shock-accelerated electrons.

The gamma-ray excess is almost certainly the IC counterpart of the
microwave haze excess described by \cite{Finkbeiner:2003im} and
\cite{Dobler:2008ww}.  Although it is still possible that a
significant fraction is prompt photons from WIMP annihilations
\cite[e.g. the 200 GeV wino advocated by][]{Grajek:2008pg} such explanations
are difficult to reconcile with the spatial similarity to the WMAP
haze (see \reffig{fermi_comp_wmap}).  The simplest hypothesis is that
the signal is mainly IC from the same electrons that produce the WMAP
haze synchrotron.

This addresses the stubborn question about the origin of the WMAP haze.
Until recently, it has been argued that the WMAP haze had 
alternative explanations, such as free--free emission from hot gas or
spinning dipole emission from rapidly rotating dust grains.  However,
the existence of this IC signal proves that the microwave haze is
indeed synchrotron emission from a hard electron spectrum.

\emph{Fermi} LAT photon data are contaminated by particle events,
especially at high energies.  We have taken care to account for the
isotropic background resulting from extragalactic sources, cosmic ray
contamination, and heavy nuclei contamination and found that this
background, though significant, is below the observed IC excess even
up to 100 GeV.  \emph{Particle contamination is extremely unlikely to
  mimic the observed signal.}

The LAT collaboration continues to refine the cuts used to define 
``diffuse class'' events, and plans to release a cleaner class of
events in coming months.  This, along with a new public version of 
GALPROP, including updated ISRF models, will allow a more sophisticated
analysis than that presented in this paper.  We eagerly await the 
release of these software and data products. 

The spectrum and morphology of both the microwave and gamma-ray haze
constrain explanations for the source of these electrons.  There have
been speculations that the microwave haze could indicate new physics,
such as the decay or annihilation of dark matter, or new astrophysics,
such as a GRB explosion, an AGN jet, or a spheroidal population of 
pulsars emitting hard electrons. 
We do not speculate in this paper on the origin of the haze electrons,
other than to make the general observation that the roughly spherical
morphology of the haze makes it difficult to explain with any
population of disk objects, such as pulsars.  The search for new
physics -- or an improved understanding of conventional astrophysics
-- will be the topic of future work.

\vskip 0.15in {\bf \noindent Acknowledgments:} We acknowledge helpful
conversations with Elliott Bloom, Jean-Marc Casandjian, Carlos Frenk, Isabel Grenier, Igor Moskalenko, Simona Murgia, Troy Porter, Andy Strong,
Kent Wood. This work was partially supported by the Director, Office
of Science, of the U.S.  Department of Energy under Contract
No. DE-AC02-05CH11231. NW is supported by NSF CAREER grant
PHY-0449818, and IC and NW are supported by DOE OJI grant \#
DE-FG02-06ER41417. IC is supported by the Mark Leslie Graduate
Assistantship.  DF and TS are partially supported by NASA grant NNX10AD85G.
TS is partially supported by a Sir Keith Murdoch
Fellowship from the American Australian Association.  This research
made use of the NASA Astrophysics Data System (ADS) and the IDL
Astronomy User's Library at Goddard.\footnote{Available at
  \texttt{http://idlastro.gsfc.nasa.gov}}

\appendix
\section{Unresolved point sources}
\label{app:pointsources}

In this section we explore the possibility that the diffuse excess
discussed in this work (the \emph{Fermi} haze) could originate from a
large number of unresolved point sources. The limit where the
contribution from unresolved point sources is dominated by emission from
many very faint ($\ll 1$ count) sources cannot be distinguished from
purely diffuse emission; however, if a smaller number of brighter ($\sim
1$ count) point sources dominate, photon events coincide within the
point spread function (PSF) at a greater rate than would be the case for
purely diffuse emission.

We model the luminosity function of unresolved point sources by a
simple power law for the expected number of counts, $S$: $dN/dS
\propto S^{-\alpha}$, with $S_\mathrm{min} <S < S_\mathrm{max}$. 
Smaller values of $\alpha$, and larger values of $S_\mathrm{max}$ and
$S_\mathrm{min}$, correspond to a larger fraction of bright point
sources and thus strengthen any upper bounds on the flux fraction from
point sources. Examining known point source populations (of AGN and X-ray
binaries) generally yields spectral indices $\alpha \sim 1.5-2.2$,
although at high luminosities steeper power laws with $\alpha \sim
2.5$ have been observed in flat spectrum radio quasars, and at low
luminosities very shallow power laws with $\alpha \sim 1.2$ have been
measured in LMXB populations \citep{Grimm:2002ta, Gilfanov:2003th, 
Kim:2003tr, Voss:2005aq, Padovani:2007qb, Abdo:2009wu}. We take $\alpha = 2.2$ and $S_\mathrm{max} = 10$, as
pessimistic benchmark parameters to provide a robust upper
bound.

This power law must break at some low-luminosity cutoff to avoid
divergences if $\alpha > 2$; we will display the statistical limits
obtained for $S_\mathrm{min} = 0.1$ and $1$ count. It is possible that any remaining excess emission could originate from point sources with expected counts/year below 0.1, but this would require the number of point sources to exceed the number of unexplained excess photons by greater than a factor of 10.

To bound the fractional flux originating from point sources, we apply a statistical
test of isotropy described in detail by \cite{Slatyer:2009zi}, designed for the case where the density of events is $\lesssim 1$ count per PSF, to the
\emph{Fermi} LAT data in the haze region ($|l| < 15$, $-30 < b < -10$). In brief, we compute (1) the fraction of ``isolated events'' with no neighbors
within some test radius $r$, and (2) the fraction of $10^6$ randomly
distributed circles of radius $r$ which contain no events (``empty
circles''). If $r$ is chosen appropriately (i.e. $r \sim$ PSF) then the ratio of these two quantities, $R\equiv$
``isolated''/``empty,'' is related to the fraction of counts arising from
unresolved point sources. If the average number of events per
circle of radius $r$, denoted $\lambda$, is greater than 1, then we
redefine ``isolated'' events to be those with fewer than $\lambda$
neighbors, and ``empty circles'' to be those containing fewer than
$\lambda$ events.  Neighbors are
found efficiently using the publicly available IDL routine
\texttt{spherematch}, and we take the test radius $r$ to be equal to
the estimated $1 \sigma$ value for the LAT PSF,
corresponding to 39\% flux containment\footnote{The
  energy-dependent PSF for the LAT is taken from \texttt{http://www-glast.slac.stanford.edu/software/IS/glast\_lat\_performance.htm}. The radius of 68\%
containment used by the \emph{Fermi} Collaboration would in the case of a Gaussian 
PSF be $r_{68} = 1.51\sigma$.} (our results are not very sensitive to this choice). We apply this test separately to the diffuse class data in each of the energy bins in Fig. \ref{fig:fermimaps}, removing events with a large zenith angle as described previously, and calibrate our results via Monte Carlo simulations. The smaller the PSF, the better the bounds (the PSF is an input to our Monte Carlo calibration); since the PSF size decreases with increasing energy, and is greater for
back-converting events compared to front-converting events, for
each energy bin we take the PSF for back-converting events at the minimum energy of the bin, in order to set robust limits.

%

\begin{figure*}[h]
   \includegraphics[width=0.6\textwidth]{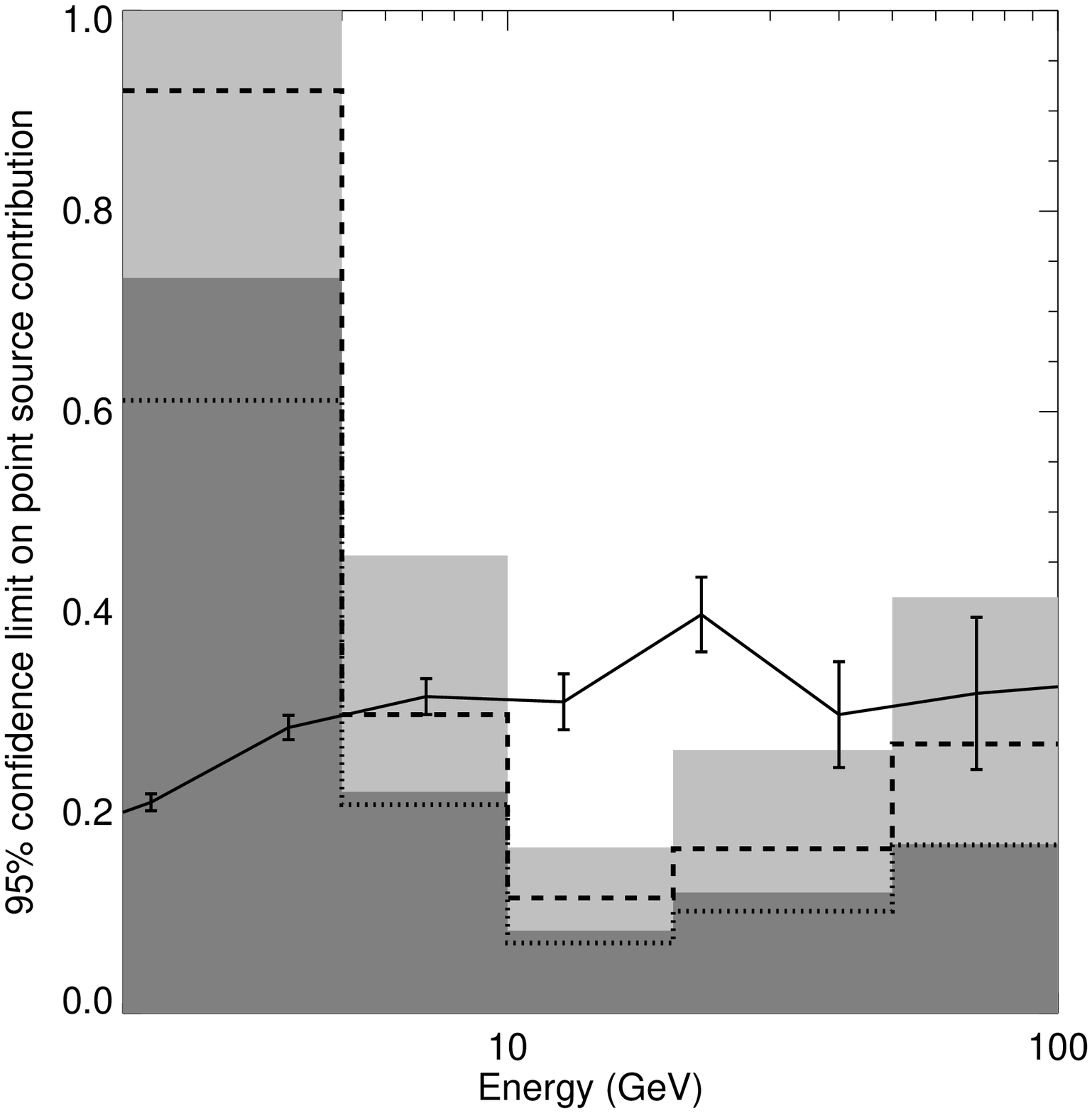}
   \includegraphics[width=0.2\textwidth]{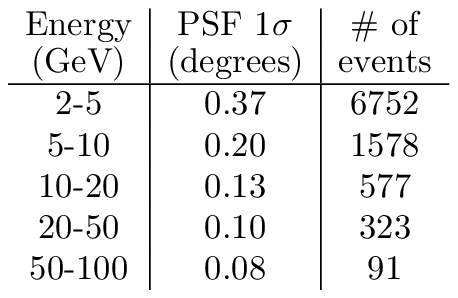}
\caption{
  Upper limits on the fraction of diffuse class events originating from unresolved point sources, in 
  the \emph{Fermi} haze region ($|l| < 15$, $-30 < b < -10$), at $95 \%$ confidence. The dark shaded region indicates bounds on sources with an expected contribution of 1+ counts/year, while the light shaded region includes sources with 0.1+ counts/year; both cases assume a luminosity function with spectral index $\alpha = 2.2$ and $S_\mathrm{max}=10$. The dotted and dashed lines indicate the bounds with a steeper luminosity function $\alpha = 1.8$, for $S_\mathrm{min} = 1$ and $S_\mathrm{min} = 0.1$ respectively. Data points indicate the fraction of emission correlated with the ``Fermi Haze'', obtained from the spectra summarized in Fig. \ref{fig:template_spec}. The
  table shows the value of $1 \sigma$ for the detector point spread function used in the Monte Carlo calibration, as conservatively estimated for
  each energy bin.
}\label{fig:pointsource} 
\end{figure*}

At low energies we find values of
$R$ consistent with a significant point source component (although the
point source fraction cannot be precisely determined without knowing
the luminosity function). Above $10$ GeV, however, the value of $R$ is
consistent with entirely diffuse emission. In the $50-100$ GeV bin, low statistics nonetheless allow the haze-correlated fraction of the emission to be explained entirely by point sources with 0.1+ expected counts/year (since in this bin the haze-correlated flux in this region corresponds to $\sim 30$ photons), but at $\sim 10-50$ GeV there is a clear excess over the $95 \%$ confidence limits on the unresolved point source contribution, for $S_\mathrm{min} = 0.1, 1$ (Fig. \ref{fig:pointsource}). There are 900 diffuse class events in this energy range and spatial region which pass the zenith angle cut, of which $\sim 1/3$ are attributed to the haze: in order for point sources providing an expected flux of less than 0.1 counts/year to make up the difference, there would have to be 1000+ such sources emitting 10-50 GeV gamma rays in this 600-square-degree region of the sky (in addition to similar source populations in the other regions where the haze-correlated emission is bright). Therefore, even for these conservative assumptions, \emph{it seems very unlikely that the hard
  spectral shape of the} Fermi \emph{haze could be due to point source contamination over the entire relevant energy range.}

\section{\emph{Fermi} full-sky map processing and data release}
\label{sec:datarelease}

In this section we describe full-sky maps generated from photon events in
the first year data release of the \emph{Fermi Gamma-Ray Space
  Telescope}.  These maps are corrected for exposure, point
source masked, and are smoothed to a Gaussian PSF, usually of
2$\degree$ FWHM.  The lowest energy maps are smoothed to 3$\degree$
or 4$\degree$.  For convenience, we provide two sets of energy bins:
12 logarithmically spaced bins for use in spectral analyses
(``specbin'') and 8 somewhat larger bins with better signal/noise for
visual inspection (``imbin'').

Although the maps were generated from a public \emph{Fermi} data
release, \emph{the maps are not an official LAT data release, and the
  procedure used to make them has not been endorsed by the LAT
  collaboration.}

\subsection{Event Selection and Binning}
The LAT is a pair-conversion telescope, in which incoming photons
strike layers of tungsten and convert to $e^+e^-$ pairs, which are
then tracked (to determine direction) on the way to a calorimeter (to
determine energy).  The first year \texttt{P6\_V3\_DIFFUSE} data file
contains records of 15,878,650 events, providing the time,
energy, arrival direction, (with respect to the spacecraft, and in
celestial coordinates), and zenith angle for each event.  Events are divided into 3
classes, using a number of cuts to separate photon events from
particle background.  ``Class 3'' rejects the largest fraction of
background contamination, and these events are used to study diffuse
emission.  We also require the zenith angle be less than $105\degree$ to 
exclude most atmospheric gammas.  By choosing to use these cuts, we study the events most
likely to be real gamma rays, at the expense of a smaller effective
area.  These cuts discard roughly
3/4 of the signal, but vastly reduce the noise.

The class 3 events are then binned in energy and into spatial pixels
to produce counts maps.  We use the hierarchical equal-area
isolatitude pixelization (HEALPix), a convenient iso-latitude
equal-area full-sky pixelization widely used in the CMB
community.\footnote{HEALPix software and documentation can be found at
  \texttt{http://healpix.jpl.nasa.gov}, and the IDL routines used in
  this analysis are available as part of the IDLUTILS product at
  \texttt{http://sdss3data.lbl.gov/software/idlutils}.}

\subsection{Exposure maps}
Because the exposure on the sky is non-uniform, we generate an
exposure map using the \texttt{gtexpcube} tool developed by the LAT
team.\footnote{
  \texttt{http://fermi.gsfc.nasa.gov/ssc/data/analysis/documentation}}
The exposure for each pixel is the LAT effective area at each $\theta$
(angle with respect to the LAT axis) summed over the livetime
of the LAT at that $\theta$, and has units of cm$^2$s.  The
exposure map spatially modulates the signal $\pm 20\%$ from raw photon
counts and is slightly energy dependent for photon energies $E > 1$
GeV.  To avoid systematic errors in generating the exposure maps we
use the same energy grid used for the spectral plots, which is given
in Table \ref{tbl:background}.  The default spatial binning of $1\degree$ in
latitude and longitude is adequate, as neighboring bins have exposure
differences of $< 0.5\%$.  The LAT contains 12 thin layers of tungsten
designated ``front'' and 4 thicker layers designated ``back.''
Photons may convert to an $\epp$ pair in any of the layers, and events
are labeled ``front'' or ``back'' accordingly.  The effective area as
a function of energy is different for front and back events, so we use
the ``\texttt{P6\_V3\_DIFFUSE::FRONT}'' and
``\texttt{P6\_V3\_DIFFUSE::BACK}'' instrumental response functions,
respectively.  Maps of counts are divided by the exposure and pixel
solid angle to produce intensity maps (cts/s/cm$^2$/sr).  The
``combined'' maps are simply an exposure-weighted linear combination
of the front and back intensity maps.  The full-sky \emph{Fermi} maps
are displayed in \reffig{fermimaps} along with the exposure map and
mask.

\subsection{Smoothing}
In our analysis we wish to compare \Fermi\ maps at different energies
to each other, and to other templates.  In order to match the PSFs
of all these maps, we smooth each map by an appropriate kernel. 
 On average, the front and back converting events have
different PSFs (by a factor of $\sim 2$), so we smooth 
the intensity maps of each to a common PSF (usually
$2\degree$) to produce ``front'' and ``back'' smoothed maps
(which can be averaged to obtain ``combined'' smoothed maps as above). 

For the PSF, we use a simple fit to the radius of 68\% flux 
containment (in degrees):
\be
r_{68} = ((c_1 E^\beta)^\alpha + c_2^\alpha)^{1/\alpha}
\label{eq:psf}
\ee
where $E$ is in GeV, $\alpha=1.2$, $\beta=-0.83$, and $(c_1,c_2)$ =
$(0.50, 0.04)$ and $(0.90, 0.09)$ for front- and back-converting
events, respectively.  This yields $r_{68} = c_1 E^\beta$ at low $E$
and $r_{68} = c_2$ at high $E$.
For simplicity, we assume the PSF is a Gaussian
so that the FWHM, $f = 2r_{50} \approx 1.56 r_{68}$.  The ``raw'' FWHM
of a counts map is then taken to be Eq. \ref{eq:psf} evaluated at
$E_{\rm mean} = \sqrt{E_0E_1}$, where $E_0 < E < E_1$ for the
events in the map.  In order to smooth to the target PSF (usually
$f_{\rm targ} = 2\degree$), a Gaussian smoothing kernel of size
$f_{\rm kern}$ is used such that 
\be
f_{\rm kern} = \sqrt{f_{\rm targ}^2
  - f_{\rm raw}^2}.
\ee
For the lowest energy maps ($E<1~\GeV$), where $f_{\rm raw}$ is large, we
take $f_{\rm targ} = 3$ or $4\degree$.

\subsection{Point Source Mask}
The point source mask (\reffig{fermimaps}) contains the 3-month
$10\sigma$ point source catalog, plus the LMC, SMC, Orion, and Cen A.
The mask radius for point sources is taken to be the 95\% containment
radius for the lowest energy event in the energy bin.  For unsmoothed
maps, both the counts map and the exposure map are multiplied by the
mask.  For smoothed maps, counts and exposure are multiplied by the
mask, then smoothed, then divided.  Pixels where the smoothed masked
exposure drops below 25\% of the mean unmasked exposure are replaced
with zeros.  In cases where the mask radius is much smaller than the
smoothing radius, the mask is ``smoothed away.''  In other cases, masked
pixels are visible in the smoothed maps.  Because of the energy
dependence of the mask radius, the effective mask changes with energy,
so care must be taken in cross comparisons between energy bins, 
\emph{especially within a few degrees of the Galactic plane.} 

\subsection{Data Release}
All LAT maps described in this section are available in the HEALPix
pixelization on the web.\footnote{ \texttt{http://fermi.skymaps.info}}
Maps of front- and back- converting events are available, as well as
the combined maps.  All three are available smoothed and unsmoothed,
and with and without point source masking, for a total of 12 maps at
each energy.  Both the imbin and specbin energy binnings are available,
for a total of 240 FITS files.  Grayscale and color jpegs are also
available to provide a quick overview.  Software used to make the maps
is available on request.

\section{Poisson Likelihood Analysis}
\label{sec:likelihood}

\bp
\centerline{
  \includegraphics[width=0.49\textwidth]{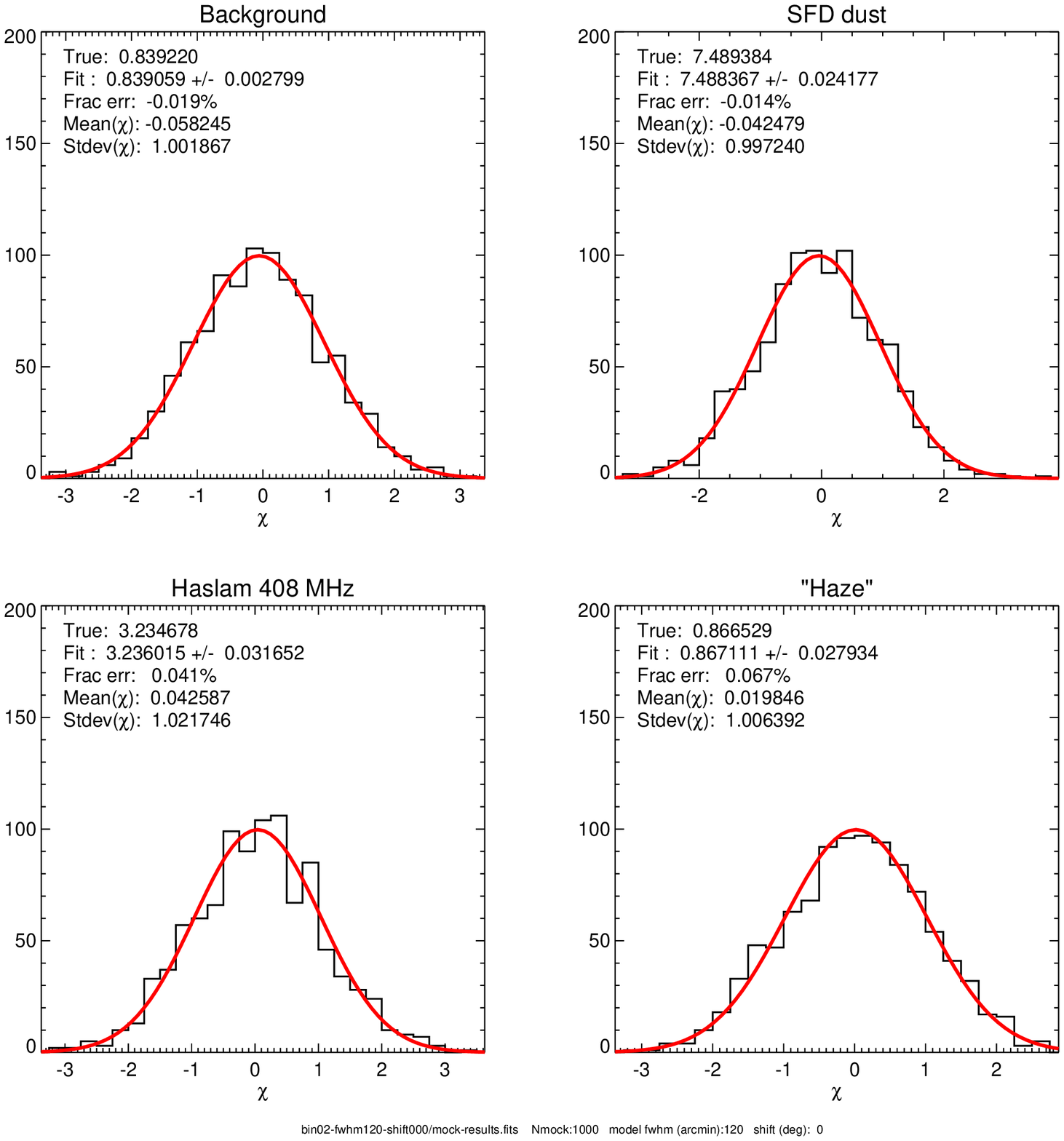}
  \includegraphics[width=0.49\textwidth]{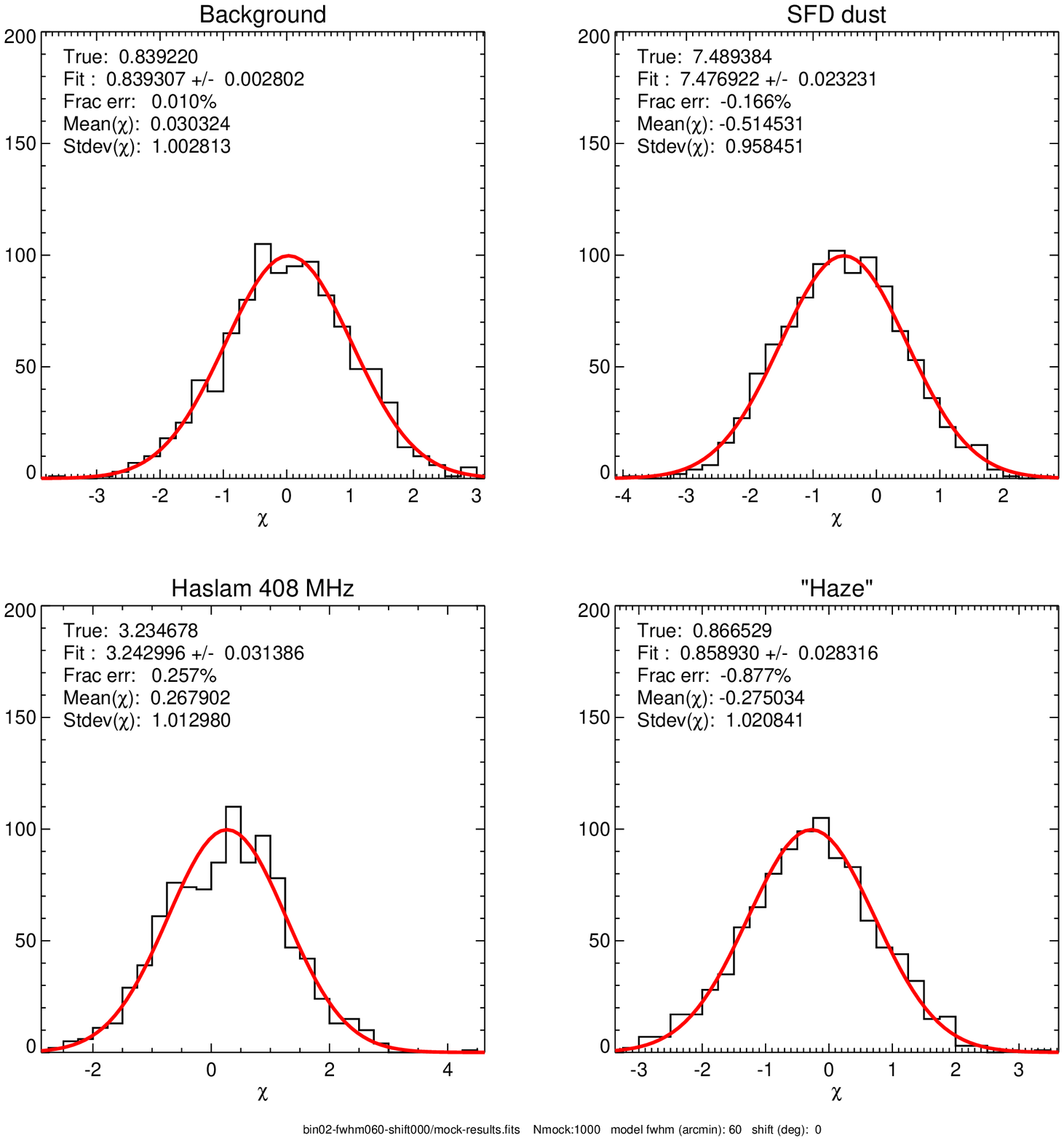}
}
\caption{ 
  \emph{Left panel:} a $2\degree$ FWHM mock map fit to a $2\degree$
  model.  \emph{Right panel:} a $1 \degree$ mock map is smoothed to a
  $2\degree$ PSF and then fit to the $2\degree$ model.  In each case,
  1000 realizations are computed, and the histograms of $\chi$ are
  shown for the four parameters.  In each case, the $\chi$ histogram
  has a mean near zero and RMS near 1, indicating that the error
  analysis is correct, even in the case of a smoothed map.  Biases
  introduced are small ($\ll 1 \sigma$) for the case shown, and are
  relatively even less important at higher energy.  }
\label{fig:mock060}
\ep

Ideally the spatial templates (dust map, synchrotron map, etc.) going in
to the Type 3 likelihood analysis (\refsec{fourcomponent}) would be
smoothed to the PSF of the gamma-ray data, allowing an event-by-event
likelihood to be calculated.  However, at some energies, the LAT has a
higher angular resolution than our templates.  In this case, the LAT
data and templates must be smoothed to a common PSF.  In this section we
explain the Poisson likelihood analysis in more detail, and check that a
Poisson likelihood evaluated on a smoothed map behaves as expected.

For the Type 3 fit we maximize the Poisson likelihood of the 4-template
model in order to weight the \Fermi\ data properly.  In other words, for
each set of model parameters, we compute the log likelihood
\be
  \ln {\mathcal L} = \sum_i \left[ k_i\ln\mu_i - \mu_i - \ln(k_i!)\right],
\ee
where $\mu_i$ is the synthetic counts map (i.e., linear combination of
templates times exposure and mask) at pixel $i$, and $k$ is the map of observed counts.  The
$\mu_i$ term depends only on the model parameters, and is not affected
by smoothing the data.  The $\ln(k_i!)$ term does not depend on the
model parameters, and so it cannot affect the best-fit model or
uncertainties.  Furthermore, there is no problem evaluating the likelihood 
for fractional $k$ using the gamma function.  
So the $k_i\ln\mu_i$ is the only potentially problematic term. 

In order to investigate the effects of smoothing, we generate each mock
map by taking a linear combination of templates (including the uniform
background), multiplying it by the exposure and mask to obtain a map of
predicted counts, and Poisson sampling it in HEALPix pixels.  This map
is then passed to our parameter estimation code to obtain the best fit
values and uncertainties for the 4 parameters. Repeating this procedure
for 1000 mock maps, we compute 
\be
\chi = \left(\frac{\mathrm{fit-true}}{\sigma} \right)
\ee
for each parameter and plot the histograms (Figure \ref{fig:mock060}).
An unbiased fit corresponds to $\langle\chi\rangle=0$ and a correct
uncertainty estimate corresponds to stdev($\chi$) = 1.  The uncertainties
appear to be correctly estimated and the bias is small.  Note that
this behavior is different from an $\exp(-\chi^2/2)$ likelihood on a
smoothed map: in that case the correlated noise induced by the smoothing
must be taken account of carefully. 

The above analysis was done for a range of energy bins with similar 
results.  The analysis was also completely redone with a $\chi^2$ 
minimization instead (indeed, this provides the initial guess for 
the Poisson likelihood minimization) with very similar results. 
Because we feel the Poisson likelihood is more correct for the 
problem at hand, we use it for the ``official'' results.

\end{document}